\documentclass[]{aa}

\usepackage{txfonts}
\usepackage{natbib,graphicx,amssymb}
\usepackage[english]{babel}
\usepackage{mathrsfs}
\usepackage{url}
\usepackage{float}

\def\revised{}

\begin{document}

\title{
The nature of the interstellar medium of the starburst low-metallicity galaxy Haro\,11: 
a multi-phase model of the infrared emission
\thanks{{\it Herschel} is an ESA space observatory with science instruments 
provided by European-led Principal Investigator consortia and with important 
participation from NASA.}
}

\author{
  D.~Cormier\inst{1}
  \and V.~Lebouteiller\inst{1}
  \and S.~C.~Madden\inst{1}
  \and N.~Abel\inst{2}
  \and S.~Hony\inst{1}
  \and F.~Galliano\inst{1} 
  \and M.~Baes\inst{3}
  \and M.~J.~Barlow\inst{6}
  \and A.~Cooray\inst{4}
  \and I.~De~Looze\inst{3}
  \and M.~Galametz\inst{5}
  \and O.~\L.~Karczewski\inst{6}
  \and T.~J.~Parkin\inst{7}
  \and A.~R{\'e}my\inst{1} 
  \and M.~Sauvage\inst{1}
  \and L.~Spinoglio\inst{8}
  \and C.~D.~Wilson\inst{7}
  \and R.~Wu\inst{1} 
}

\institute{
  Laboratoire AIM, CEA/DSM - CNRS - Universit\'e Paris
  Diderot, Irfu/Service d'Astrophysique, CEA Saclay, 91191
  Gif-sur-Yvette, France \email{diane.cormier@cea.fr} 
  \and
 University of Cincinnati, Clermont College, Batavia, OH, 45103, USA
 \and 
 Sterrenkundig Observatorium, Universiteit Gent, Krijgslaan 281 S9, B-9000 Gent, Belgium
 \and
 Department of Physics \& Astronomy, University of California, Irvine, CA 92697, USA
 \and
 Institute of Astronomy, University of Cambridge, Madingley Road, Cambridge, CB3 0HA, UK
 \and
 Department of Physics and Astronomy, University College London, Gower Street, London WC1E 6BT, UK
 \and
 Dept. of Physics \& Astronomy, McMaster University, Hamilton, Ontario, L8S 4M1, Canada
 \and
 Istituto di Fisica dello Spazio Interplanetario, INAF, Via del Fosso del Cavaliere 100, I-00133 Roma, Italy
 }


\abstract
{The low metallicity interstellar medium (ISM) is profoundly different from that of normal systems, 
being clumpy with low dust abundance and little CO-traced molecular gas. 
Yet many dwarf galaxies in the nearby universe are actively forming stars. 
As the complex ISM phases are spatially mixed with each other, detailed modeling 
is needed to understand the gas emission and subsequent composition and structure 
of the ISM.}
{Our goal is to describe the multi-phase ISM of the infrared bright low-metallicity 
galaxy Haro\,11, dissecting the photoionised and photodissociated gas components.}
{We present observations of the mid-infrared and far-infrared fine-structure 
cooling lines obtained with the {\it Spitzer}/IRS and {\it Herschel}/PACS spectrometers.
We use the spectral synthesis code Cloudy to methodically model 
the ionised and neutral gas from which these lines originate.}
{We find that the mid- and far-infrared lines account for $\sim$1\% of the total 
infrared luminosity $\rm{L_{TIR}}$, acting as major coolants of the gas. 
Haro\,11 is undergoing a phase of intense star formation, as traced by 
the brightest line, [O~{\sc iii}]~88~$\mu$m, with $\rm{L_{[O~III]}/L_{TIR}}$~$\sim$~0.3\%, 
and high ratios of [Ne~{\sc iii}]/[Ne~{\sc ii}] and [S~{\sc iv}]/[S~{\sc iii}].  
Due to their different origins, the observed lines require a multi-phase modeling comprising: 
a compact H~{\sc ii} region, dense fragmented 
photodissociation regions (PDRs), 
a diffuse extended low-ionisation/neutral gas which has a 
volume filling factor of at least 90\%, 
and porous warm dust in proximity to the stellar source. 
For a more realistic picture of the ISM of Haro\,11 we would need to model 
the clumpy source and gas structures. 
We combine these 4 model components to explain the emission of 17 
spectral lines, investigate the global energy balance of the galaxy through 
its spectral energy distribution, and establish a phase mass inventory.
While the ionic emission lines of Haro\,11 essentially originate from 
the dense H~{\sc ii} region component, a diffuse low-ionisation gas is needed 
to explain the [Ne~{\sc ii}], [N~{\sc ii}], and [C~{\sc ii}] line intensities. 
The [O~{\sc iii}]~88~$\mu$m line intensity is not fully reproduced by our model, hinting 
towards the possible presence of yet another low-density high-ionisation medium. 
The [O~{\sc i}] emission is consistent with a dense PDR of low covering factor, 
and we find no evidence for an X-ray dominated component.
The PDR component accounts for only 10\% of the [C~{\sc ii}] emission. 
Magnetic fields, known to be strong in star-forming regions, may dominate the 
pressure in the PDR. For example, for field strengths of the order of 100~$\mu$G, 
up to 50\% of the [C~{\sc ii}] emission may come from the PDR.
}
{}

\titlerunning{Modeling the ISM of Haro\,11}
\authorrunning{Cormier et al.}
\maketitle

\section{Introduction}
Star formation in the most pristine environments of the early universe 
is poorly understood. The closest analogs to chemically unevolved 
systems are the low metallicity dwarf galaxies of our local universe. 
But even those well-studied nearby galaxies continue to plague
us with fundamental questions on what triggers and regulates star formation 
in metal-poor gas-rich conditions. 
Certainly, from an observational point of view, at all wavelengths, 
local universe low-metallicity dwarf galaxies show dramatic differences 
compared to their more metal-rich counterparts \citep[][for a review]{kunth-2000}, 
in the mid-infrared (MIR), far-infrared (FIR), and 
submillimeter (submm) wavelength regimes. 
These include the dearth of Polycyclic Aromatic Hydrocarbons (PAHs), 
prominent hot MIR-emitting dust, submm excess, and enhanced 
[C~{\sc ii}]~157~$\mu$m/CO(1-0) ratios \citep[e.g.][]{madden-2008,madden-2012a}. 

What physical properties, local as well as global, within the dwarf galaxies 
are we witnessing with these observational signatures? How do we turn 
these signatures into a realistic view of star formation under early universe conditions? 
How do basic local parameters, such as radiation field, density, compactness, filling factors, 
metallicity, etc. figure in the picture we have of star-forming dwarf galaxies? 
Much of the ambiguity surrounding these questions is due to the lack of 
understanding the precise role of critical diagnostics and of spatial resolution 
in the most nearby objects, especially in the FIR/submm, resulting in a mixture 
of different interstellar medium (ISM) phases in the integrated view of galaxies.  

Low metallicity star-forming dwarf galaxies host young massive stellar clusters 
\citep[e.g.][]{turner-2000,beck-2000,johnson-2003} which produce ultraviolet (UV) photons. 
\revised{
The low abundance and prevailing hard radiation field can explain most of the differences 
observed in dwarf galaxies in destroying large grains and PAHs, in favor of smaller 
dust grains. The UV photons also photodissociate molecules such as CO, which 
are unable to adequately self-shielding, thus reducing the CO abundance in metal-poor gas 
\citep[e.g.][]{wolfire-2010,narayanan-2012}. 
}
Thus UV photons play an important role in the heating 
and the chemistry of the ISM, controlling the structure of H~{\sc ii} regions and photodissociation 
regions (PDRs). While the heating of the gas is mainly due to photoionisation and gas-grain collisions 
in the H~{\sc ii} regions and to the photoelectric effect and cosmic rays in the neutral phase, 
the cooling is normally dominated by radiative de-excitation of fine-structure lines, 
readily observed in the MIR and FIR/submm. 
The behavior of these tracers and their relationship to the star formation properties, 
have been extensively studied to access physical parameters in terms of the gas temperature, 
density, and the radiation field \citep[e.g.]{malhotra-2001,hunter-2001,brauher-2008}. 
These sudies make use of photoionisation and PDR models at the scale of individual clouds
\citep{kaufman-1999,wolfire-1990,ferland-1998,abel-2005,roellig-2006,tielens-hollenbach-1985}.

The [C~{\sc ii}] 157~$\mu$m line is an important FIR diagnostic often put forward as 
a star formation tracer \citep{stacey-2010,boselli-2002,delooze-2011}. 
It has been observed in many different object types, including dense star-forming regions 
\citep{stacey-1991}, diffuse ionised and atomic regions \citep{madden-1993}, 
and high-redshift galaxies \citep{maiolino-2009,stacey-2010}. 
Since the [C~{\sc ii}] line is well-correlated with the $^{12}$CO(1-0) line and 
the FIR luminosity \cite[e.g.][]{stacey-1991,pierini-2003}, it is generally considered that 
the [C~{\sc ii}] line traces massive star formation within the PDR paradigm. 
[C~{\sc ii}] is normally the brightest FIR cooling line, 
being much brighter than CO, and carrying about 0.1 to $\sim$0.5\% 
of the FIR luminosity in metal-rich star-forming galaxies.
In low metallicity dwarf galaxies, where CO(1-0) is not easily detected 
\revised{
because of photodissociation 
}
\citep[e.g.][]{schruba-2012,leroy-2011}, [C~{\sc ii}] can be as high as 1\% of the 
L$\rm{_{FIR}}$ and often 5 to 10 times brighter relative to CO than in dustier starburst galaxies 
\citep[e.g.][]{poglitsch-1995,madden-1995,madden-2000,hunter-2001,cormier-2010}. 
The high [C~{\sc ii}]/CO(1-0) ratio in low-metallicity systems may be alerting us to 
the fact that the morphology of their PDRs and H~{\sc ii} regions and the relative 
filling factors of the various phases of dwarf galaxies differ from those of more 
metal-rich objects \citep[e.g.][]{madden-2000,kaufman-2006,roellig-2006}. 
Even in the nearest neighboring dwarf galaxies, the picture of star formation 
-- and what controls this star formation -- is enigmatic.

However, using [C~{\sc ii}] alone can lead to ambiguous interpretation, 
particularly on the origin of the C$^+$ emission, which 
can arise from the ionised as well as neutral phases of the ISM. 
To employ the [C~{\sc ii}] line as a valuable star formation tracer, 
it is necessary to model the different phases on galaxy-wide scales. 
The challenge rests on the unavoidable fact that on the scale of galaxies, 
aside from the closest dwarf galaxies such as the LMC (50~kpc, 1/2~Z$_{\odot}$), 
SMC (60~kpc, 1/5~Z$_{\odot}$), NGC\,6822 (500~kpc, 1/4~Z$_{\odot}$), 
or IC\,10 (800~kpc, 1/3~Z$_{\odot}$), 
the phases of the ISM are mixed together in single telescope beams. 
When using a limited number of diagnostic lines, 
only a single galactic component can be modeled and average conditions derived, 
which is otherwise an ensemble of individual PDRs, ionised regions, etc. 
This requires modeling a comprehensive suit of tracers covering a range 
of critical densities and excitation potentials to characterise the dense 
H~{\sc ii} regions, the diffuse H~{\sc ii} regions, PDRs, etc. self-consistently.  
This approach allow us to grapple with degeneracies and ambiguities in the interpretation 
of these tracers. In this way, valuable observational diagnostics can be turned 
into a more accurate description of the multi-phase configuration of a full galaxy. 

The {\it Herschel} Space Observatory \citep{pilbratt-2010} 
has brought the opportunity to obtain 
a wide range of valuable FIR fine-structure diagnostics never observed 
before in dwarf galaxies. The Dwarf Galaxy Survey (DGS; P.I. Madden) 
is surveying the dust and gas of a wide range of low metallicity 
galaxies in the local universe \citep{cormier-2010,galametz-2010,o'halloran-2010,madden-2012b,remy-2012} 
using PACS \citep{poglitsch-2010} and SPIRE \citep{griffin-2010}. 
The brightest low-metallicity galaxy of the DGS observed with the 
PACS spectrometer is \object{Haro\,11}, an infrared-luminous starburst 
galaxy with metallicity $\rm{Z=1/3~Z_{\odot}}$. 
Haro\,11 has the largest dataset obtained with PACS 
with observations of 7 FIR fine-structure lines: [C~{\sc ii}]~157, [O~{\sc iii}]~88, 
[O~{\sc i}]~63, [O~{\sc i}]~145, [N~{\sc iii}]~57, [N~{\sc ii}]~122, and 
[N~{\sc ii}]~205~$\mu$m. 
It has also been observed with the IRS spectrograph onboard {\it Spitzer}. 
Unresolved in the IR, Haro\,11 is an ideal candidate to conduct a 
study modeling an exhaustive set of gas tracers 
to retreive information about the structure of the multi-phase low metallicity ISM.

In this paper, we propose a systematic approach to model the multi-phase 
ISM of Haro\,11, focusing on the gas properties of the ionised and neutral ISM phases. 
We consider a set of 17 {\it Herschel} and {\it Spitzer} MIR and FIR ionic and 
atomic fine-structure cooling lines which trace a wide range of environmental 
conditions, spanning excitation potentials from 8 to 55~eV. 
With the spectral synthesis code Cloudy \citep{ferland-1998}, we create a 
self-consistent model of the dominant ISM components from which these lines originate
in order to understand the prevailing mechanisms and physical conditions.  
In Section~\ref{sect:obs}, we present spectroscopic data from the {\it Spitzer} and 
{\it Herschel} observatories. Section~\ref{sect:model} explains the modeling strategy 
and assumed initial parameters. Section~\ref{sect:results},~\ref{sect:diffus}, and~\ref{sect:disc} 
report the model results and conditions of the ISM components. 
This study establishes an encompassing picture of the dominant ISM phases 
in Haro\,11 and describes the mass budget of the various phases in our model 
(Section~\ref{sect:buildup}).

\subsection{Studies of Haro\,11}
\label{sect:haro11}
\begin{figure}[!htp]
\centering
\includegraphics[clip,trim=2cm .5cm 2cm 2cm,width=8.8cm]{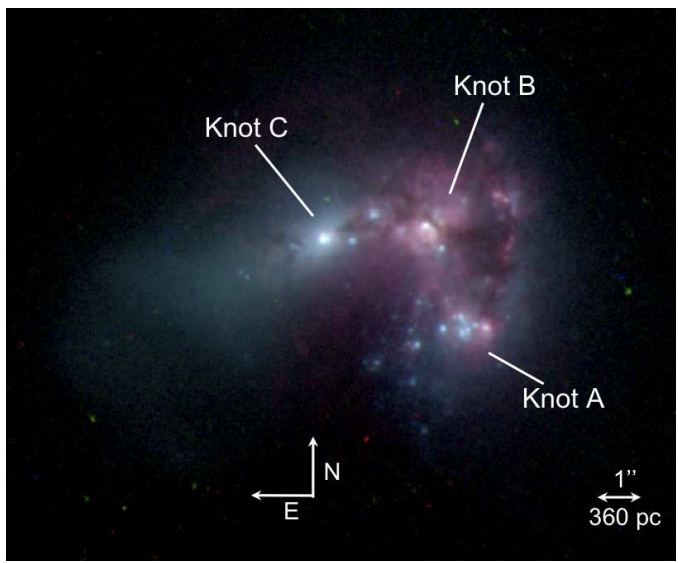}
\caption{
HST/ACS 3-color image of Haro\,11 
with labels for the 3 star-forming knots
\revised{
(blue: F435, green: F550, red: FR656N/H$\alpha$). 
}
Images were downloaded from the Hubble Legacy Archive.
}
\label{fig:acs}
\end{figure}

Several studies of Haro\,11 have highlighted its peculiarity as a galaxy 
and challenged our understanding of star formation. 
Haro\,11 -- also know as ESO~350-IG38 -- 
is a blue compact dwarf (BCD) galaxy at 84~Mpc, with M$\rm{_{B} =-20.3}$, 
and metallicity Z$\sim$1/3~Z$_{\odot}$ \citep{guseva-2012}. 
It is composed of 3 main star-forming regions, or knots, described in 
\cite{vader-1993} and \cite{kunth-2003}, with a morphology presumably 
resulting from a merger event \citep{ostlin-2001}. 
The starbursting nature of Haro\,11 is visible in Figure~\ref{fig:acs} with 
a dominant bright stellar component and dust lanes in front of knot B.
These knots host hundreds of young star clusters and $\sim$60 super 
star clusters (SSC) with ages between 1 and 100 Myr peaking at 3 Myr, 
and constituting a total stellar mass of $\rm{10^{10}~M_{\odot}}$ 
\citep{adamo-2010}.
An older stellar population is also present, primarily in knots A and C, 
as seen in the red colors of the V-K bands, modeled in \cite{micheva-2010} 
by a metal-poor (Z=0.001) stellar population of age around 14~Gyr with 
standard Salpeter IMF, indicating that the galaxy is not undergoing its first star formation event.  
Ly$\alpha$ emission \citep{kunth-1998} and Lyman continuum leakage were investigated 
using UV and X-ray observations in \cite{bergvall-2006,hayes-2007,grimes-2007} 
and \cite{leitet-2011}. In \cite{leitet-2011}, they estimated the leakage to be 3\%.
In particular, the discrepant location of Ly$\alpha$ and H$\alpha$ emission, originating 
in the ionised gas surrounding the star-forming regions, may be explained by 
a diffuse ionised ISM in the halo of the galaxy with internal resonant scattering 
at the surface of clumpy dense neutral clouds \citep{kunth-2003,hayes-2007}.
These properties have resulted in its popularity as a local analogue 
of the high-redshift Lyman-break galaxies \citep{overzier-2008}, 
and therefore a must-study case to link nearby star formation to the 
distant universe and galaxy evolution.

In the IR, Haro\,11 is extremely bright compared to other nearby dwarf galaxies. 
IRAS measurements yielded a FIR luminosity L$\rm{_{FIR}}$ of 
1~$\times$~10$^{11}$~L$\rm{_{\odot}}$ \citep{sanders-2003} from 40 to 400~$\mu$m 
using the \cite{lonsdale-1985} prescription.
Its spectral energy distribution (SED) shows a peculiar behavior. 
It peaks around 40~$\mu$m, indicating the prevalence of warm dust, 
and shows an excess of emission in the submm, as unveiled by {\it APEX}/LABOCA 
and {\it Herschel}, which is often seen in low metallicity starbursting dwarf galaxies 
\citep{galliano-2003,galliano-2005,galametz-2009,galametz-2011,remy-2012}. 
The origin of the excess is not well determined and several hypotheses 
described in \cite{galametz-2011}, including a change of dust emissivity, 
spinning dust grains, or a cold dust component, are invoked in 
current dust models to explain the submm excess. 
\cite{galametz-2009} present a detailed dust modeling of Haro\,11
in which they derive a total dust mass of $\rm{6 \times 10^{6}~M_{\odot}}$ 
if the submm excess is ignored, and of $\rm{2 \times 10^{7}~M_{\odot}}$ 
if the submm excess is interpreted as cold dust. 
Assuming the excess as cold dust (T$\sim$10~K) is however 
difficult to reconcile with the observed dust and gas mass budgets, 
and the expected lower dust-to-gas mass ratios \citep[DGR;][]{galametz-2011}. 
Integrating the SED from 3 to 1100~$\mu$m, the total IR luminosity L$\rm{_{TIR}}$ is 
1.4~$\times$~10$^{11}$ L$\rm{_{\odot}}$ \citep{galametz-2009}. 
We will use this value of L$\rm{_{TIR}}$ throughout this paper.

Spectroscopic observations show bright lines in the FIR with {\it ISO}/LWS, 
while both the H~{\sc i} 21-cm line, and the CO~(J=1-0) line, the most common 
tracer of molecular gas in galaxies, are undetected \citep{bergvall-2000}. 
A detailed study of the neutral gas of the ISM of Haro\,11 by \cite{bergvall-2000} 
investigated the mass budget of the different ISM phases. 
They establish masses for each phase: 
M(H~{\sc i}) $\rm{< 10^{8}~M_{\odot}}$, with an upper limit on the atomic H~{\sc i}; 
M(PDR) of $\rm{2~\times~10^{8}~M_{\odot}}$ using the LWS [C~{\sc ii}] observations, 
the PDR mass being larger than the H~{\sc i} mass; 
and $\rm{M(H_{2}) \sim 10^{8}~M_{\odot}}$ for the molecular gas, using both a scaling 
of the PDR mass and the CO upper limit converted to H$\rm{_{2}}$ 
column density through the CO/H$\rm{_{2}}$ conversion factor ($X_{\rm CO}$). 
They used a Galactic value for $X_{\rm CO}$ \citep{strong-1988}: 
$X_{\rm CO, gal}=2.3 \times 10^{20}~\rm{cm^{-2}~(K~km~s^{-1})^{-1}}$. 
Therefore they describe the ISM structure as: very extended PDRs, in which the 
H~{\sc i} mass resides; a small fraction of the total ISM gas is left in atomic form; 
and the bulk of the gas, that is normally in neutral state, is ionised by the starburst.
\cite{bergvall-2002} derive M(H~{\sc ii}) $\rm{\sim 10^{9}~M_{\odot}}$ 
from H$\alpha$ observations for the ionised gas. 
These masses are lower than the total stellar mass of $\rm{10^{10}~M_{\odot}}$ 
derived in \cite{adamo-2010}. 

However, important questions remain, 
particularly on the state of the ionised, neutral atomic, and molecular gas. 
While evidence for young massive stellar populations and SSCs exist from the 
optical and IR point of view, evidence for any substantial reservoir of neutral gas, 
in the form of either atomic or molecular, is non-existent to date. 
Where is the fuel for the vigorous star formation observed in Haro\,11? 
What is the state of the gas reservoirs in Haro\,11, residing in the ionised, neutral, 
and molecular form?
This study, using a wide variety of tracers of neutral and ionised gas, aims for 
a more complete description of the enigmatic nature of the ISM of Haro\,11.

\begin{center}
\begin{table}[!htp]
  \caption{General properties of Haro\,11.}
  \begin{tabular}{l l l}
    \hline\hline
    \multicolumn{1}{l}{Quantity} & 
    \multicolumn{1}{l}{Value} &
    \multicolumn{1}{l}{Reference} \\
    \hline
	RA (J2000)	& 00h36$^{\prime}$52.5$^{\prime\prime}$	& NED $^{(a)}$ 	\\
	Dec (J2000)	& -33$^{\circ}$33$^{\prime}$19$^{\prime\prime}$	& NED \\
	Distance	& 84~Mpc	& NED $^{(b)}$ \\
	12+log(O/H)	& 8.2		& \cite{guseva-2012}	\\
	Optical size	& 0.4$^{\prime}$x0.5$^{\prime}$	& NED	\\
	$L_{\rm TIR}$$^{(c)}$	& 1.4 $\times \rm{10^{11}~L_{\odot}}$	& \cite{galametz-2009}	\\ 
	$L_{\rm FUV}$	& 2 $\times \rm{10^{10}~L_{\odot}}$		& \cite{grimes-2007}	\\
	$L_{\rm X-rays}$	& 4.4 $\times \rm{10^{7}~L_{\odot}}$	& \cite{grimes-2007}	\\
	$L_{\rm Ly\alpha}$	& 1.9 $\times \rm{10^{8}~L_{\odot}}$	& \cite{hayes-2007}	\\ 
	$L_{\rm H\alpha}$	& 8 $\times \rm{10^{8}~L_{\odot}}$	& \cite{ostlin-1999}	\\ 
    \hline \hline
  \end{tabular}
  \begin{flushleft}
  $(a)$ NASA/IPAC Extragalactic Database.~
  $(b)$ This distance is derived with a Hubble constant 
  H$_0$~=~73~km/sec/Mpc.~
  $(c)$ L$\rm{_{TIR}}$ is the integrated SED from 3 to 1100~$\mu$m.~
  All luminosities are scaled to the distance 
  chosen in this paper of 84~Mpc. 
  \end{flushleft}
  \label{table:general}
\end{table}
\end{center}

\section{Observations}
\label{sect:obs}
\subsection{{\it Spitzer} data}
\label{sect:irs}
\begin{figure}[!tp]
\centering
\includegraphics[clip,trim=.5cm 0 1cm 0,width=8.8cm]{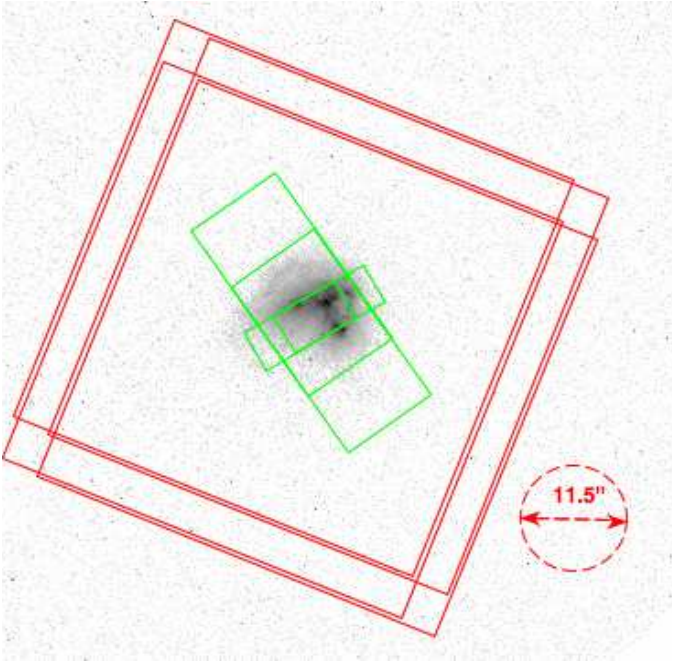}
\caption{H$\alpha$ image of Haro\,11 from the Hubble Legacy Archive. 
The {\it Spitzer}/IRS high-resolution slits are overlaid in green and 
the {\it Herschel}/PACS array in red.
The PACS beamsize at 150~$\mu$m is 11.5$^{\prime\prime}$. 
The IRS low-resolution slits are not shown.
}
\label{fig:aors}
\end{figure}

Haro\,11 was observed with the Infrared Spectrograph \citep[IRS;][]{houck-2004} 
on board the {\it Spitzer} Space Telescope \citep{werner-2004}, 
on the 17th of July 2004 (P.I.: J. R. Houck). 
The 4 modules were used, i.e., the 2 high-resolution modules: 
Short-High SH ($\approx10-20~\mu$m) and Long-High LH ($\approx19-37~\mu$m) 
with $R\sim600$, and the 2 low-resolution modules: Short-Low SL ($\approx5-14~\mu$m) 
and Long-Low LL ($\approx14-37~\mu$m) with $R\sim60-120$. 
Observations were done in staring mode (AORKey 9007104). 
Figure~\ref{fig:aors} shows the slits of the high-resolution module on the sky.

\begin{figure*}[!htp]
\centering
\includegraphics[width=17cm]{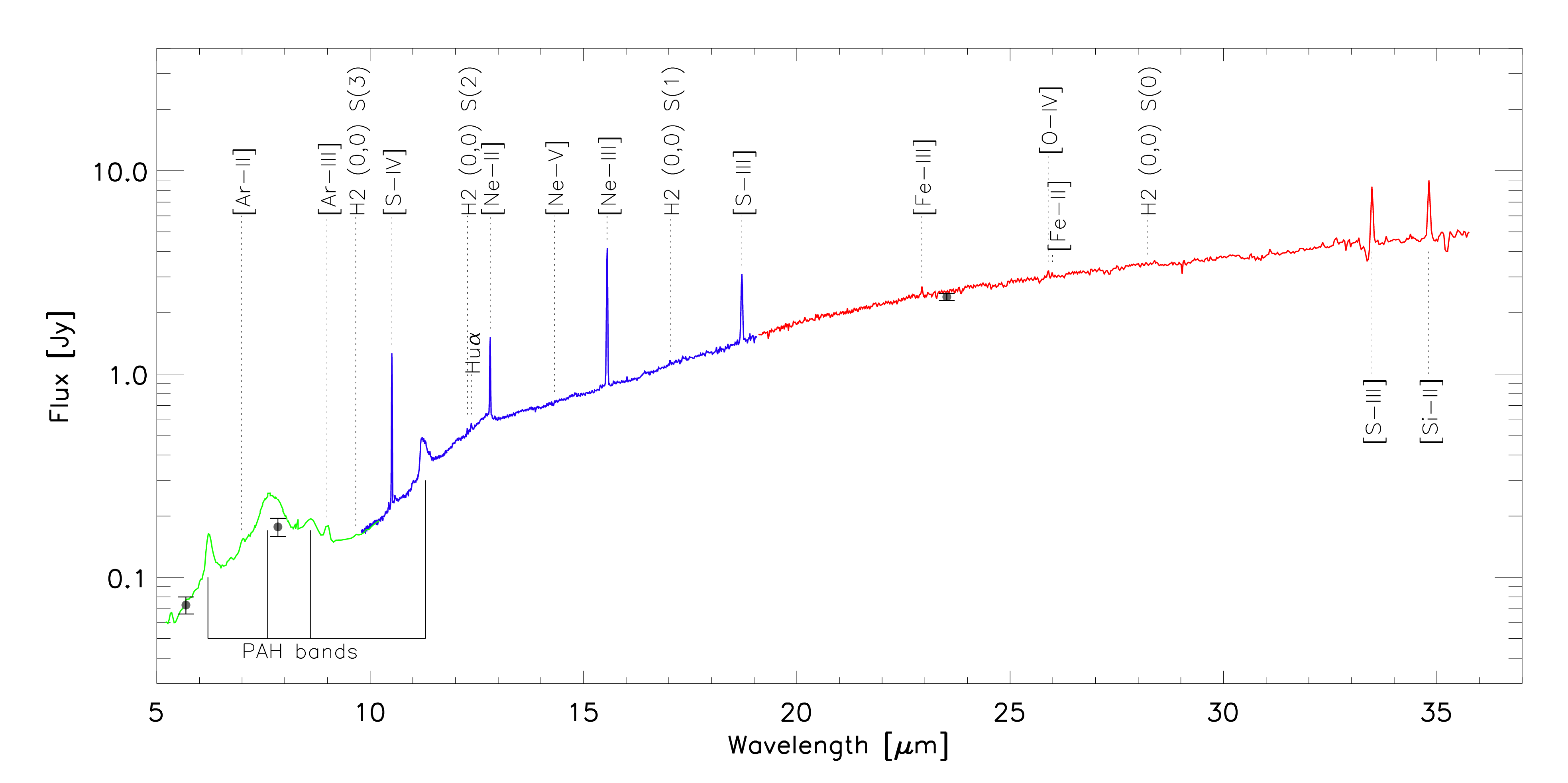}
\includegraphics[width=17cm]{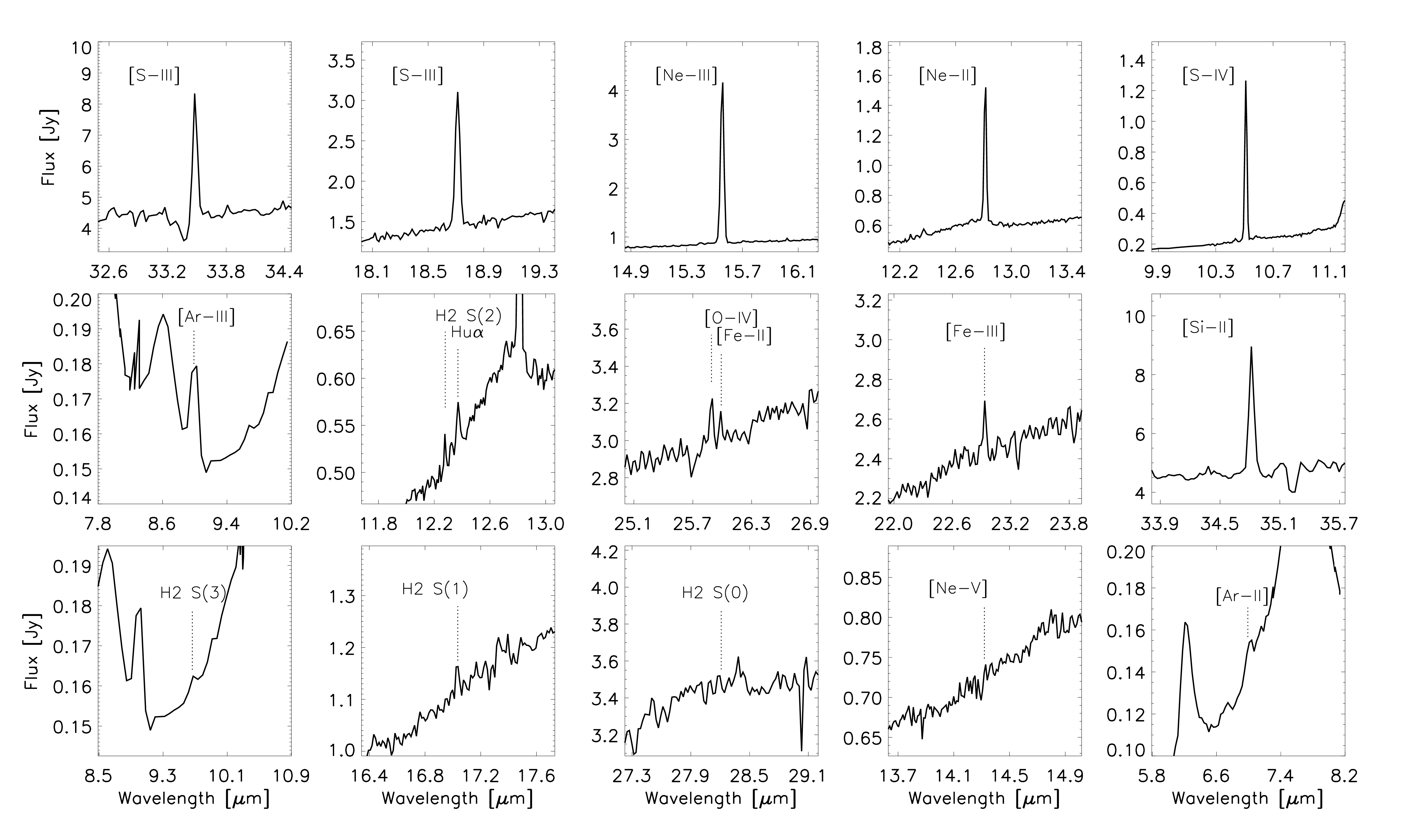}
\caption{ 
\textit{Top:} 
{\it Spitzer}/IRS spectrum of Haro\,11 
in the rest wavelength frame ($\rm{z=0.0206}$), 
SL in green, SH in blue, and LH in red. 
Emission lines are labeled on top. 
The IRAC and MIPS photometry is also overlaid 
(fluxes in Table~\ref{table:phot}).
\textit{Bottom:} 
Zoom on the individual spectral lines. 
Fluxes and upper limits are reported in
Table~\ref{table:lines}.
}
\label{fig:irs_spec}
\end{figure*}

The low-resolution spectra were retrieved from the CASSIS Atlas\footnote{
The Cornell Atlas of {\it Spitzer}/IRS Sources (CASSIS) is a product of the 
Infrared Science Center at Cornell University, supported by NASA and JPL.}
\citep{lebouteiller-2011}, with a tapered spatial window that scales 
with wavelength to account for the broader PSF at longer wavelengths. 
The high-resolution spectra were obtained with SMART v8.2.2
\citep{lebouteiller-2010,higdon-2004} from the full-slit extraction of SH and LH 
and assuming a point-like source calibration. The SL and SH spectra were stitched 
to LL and LH by applying the same factor $1.13$, which is due to the fact that 
the source is slightly extended for the SL and SH apertures ($3.7^{\prime\prime}$ 
and $4.7^{\prime\prime}$ height respectively), 
while it is point-like for the LL and LH apertures ($\approx11^{\prime\prime}$). 
Another factor of $1.15$ was applied to the low-resolution spectra 
to match the continuum of the high-resolution.
The final rest-frame spectrum is displayed in Figure~\ref{fig:irs_spec} ({\it top}), 
with a zoom on individual spectral lines ({\it bottom}).
With a spectral resolution $\ge$500~km~s$^{-1}$, the lines are not spectrally resolved.

Line fitting and flux measurements were also done in SMART. 
A Gaussian and second order polynomial function were fitted to the line and continuum. 
For total uncertainties, we added 10\% of the line flux due to calibration uncertainties. 
The resulting fluxes and uncertainties of the most prominent lines are listed in Table~\ref{table:lines}. 
The fluxes agree within the uncertainties to those of \cite{wu-2006}. 
The [Ne~{\sc iii}] line at 36.0~$\mu$m was not observed 
as it falls out of the IRS spectral range.

\subsection{{\it Herschel} data}
\label{sect:pacs}
Haro\,11 was observed with the {PACS} spectrometer \citep{poglitsch-2010} 
on the 27th of June 2010, OD 409 of the {\it Herschel} mission \citep{pilbratt-2010} 
as part of the Guaranteed Time Key Program, the Dwarf Galaxy Survey (P.I.: S. C. Madden). 
The {PACS} array is composed of 5x5 spatial pixels each 9.4$^{\prime\prime}$ square, 
covering a total field-of-view of 47$^{\prime\prime}$. 
Haro\,11 was mapped with 2x2 rasters separated by 4.5$^{\prime\prime}$ in 
the 7 following fine-structure spectral lines: 
[C~{\sc ii}] 157~$\mu$m, [O~{\sc iii}] 88~$\mu$m, [O~{\sc i}] 145~$\mu$m 
(OBSID 1342199236), [O~{\sc i}] 63~$\mu$m, [N~{\sc iii}] 57~$\mu$m, 
[N~{\sc ii}] 122~$\mu$m (OBSID 1342199237), and [N~{\sc ii}] 205~$\mu$m 
(OBSID 1342199238), for a total of 7.2h. 
The [N~{\sc ii}] 205~$\mu$m line was also re-observed on OD 942 
for 1.6h with a single pointing (OBSID 1342234063).
The observations were done in chop-nod mode
with a chop throw of 6$^{\prime}$ off the source. 
The chopped position is free of emission. 
The beam size is 9.5$^{\prime\prime}$ at 60~$\mu$m, and 
11.5$^{\prime\prime}$ at 150~$\mu$m 
(diffraction limited above $\sim$100~$\mu$m).
The spectral resolution is $\sim$~90~km~s$^{-1}$ at 60~$\mu$m, 
125~km~s$^{-1}$ at 90~$\mu$m, 295~km~s$^{-1}$ at 120~$\mu$m, 
and 250~km~s$^{-1}$ at 150~$\mu$m (PACS Observer's Manual 2011
\footnote{available at \url{http://herschel.esac.esa.int/Docs/PACS/html/pacs_om.html}}).

The data were reduced with the {\it Herschel} Interactive Processing Environment 
(HIPE) v7.0.1935 \citep{ott-2010}. 
We used standard scripts of the {PACS} spectrometer pipeline. 
The line fitting and map making were done using the software PACSman \citep{lebouteiller-2012}. 
The signal from each spatial position of the PACS array is 
fitted with a second order polynomial plus Gaussian for the baseline and line. 
For total uncertainties, we add 30\% of the line flux due to calibration uncertainties \citep{poglitsch-2010}.
The rasters are combined by drizzling to produce final reduced maps 
with pixel size of 3$^{\prime\prime}$.

\begin{figure*}[!htp]
\centering
\includegraphics[clip,width=18cm]{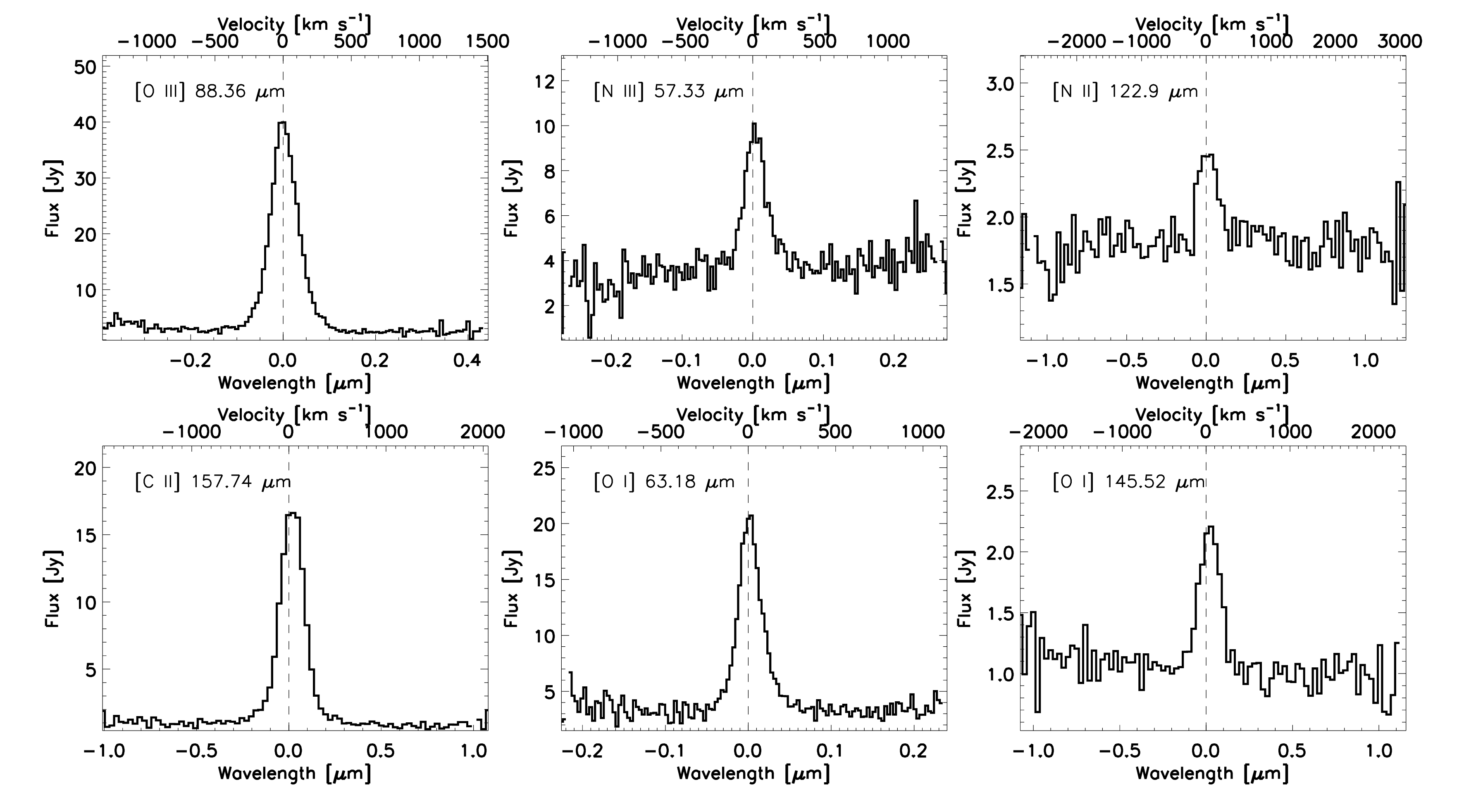}
\caption{
{\it Herschel}/PACS line profiles from an individual 
spatial pixel (9.4$^{\prime\prime}$) centered on the galaxy. 
}
\label{fig:pacs_lines}
\end{figure*}

With {\it Herschel}, all of the lines are well detected, at least in the central pixel 
of the PACS array, except the [N~{\sc ii}] 205~$\mu$m line. 
The [N~{\sc ii}] 205~$\mu$m line is located at the edge of the first grating order 
of the spectrometer and affected by spectral leakage, due to overlapping of the grating orders. 
It is contaminated by signal from the second grating order around 100~$\mu$m, and 
is therefore difficult to observe and to calibrate. 
The line profiles are displayed in Figure~\ref{fig:pacs_lines}. 
The line centers range between -30 and +70~km~s$^{-1}$ within the maps of all observed lines, 
and indicate a rotation around the north-south axis of both the ionised and neutral gas components, 
slightly tilted compared to the velocity structure of H${\rm{\alpha}}$ analysed in \cite{ostlin-1999}. 
All lines display broad profiles, with a broadening up to 300~km~s$^{-1}$, 
which suggests that both the neutral and ionised gas are coming from several 
components with different velocities. 
However, with spectral resolution $>$100~km~s$^{-1}$, the lines are well fitted by 
a single Gaussian and we cannot separate the different velocity components in the spectra.
This suggested complex velocity structure is corroborated by studies of optical lines 
that show broad profiles as well \citep{kunth-1998,ostlin-1999}, and where the neutral gas 
is shifted compared to the ionised gas, expanding over few kpcs \citep{kunth-2003}. 
FUSE spectra in \cite{grimes-2007} show absorption lines of the ISM with FWHM 
300~km~s$^{-1}$ and blueshifted by 80~km~s$^{-1}$. The higher ionisation lines 
can be stronger in galactic outflows \citep{contursi-2012}.

The [C~{\sc ii}] line shows a marginally extended emission at 157~$\mu$m, 
peaking at a signal-to-noise ratio of 50 in the center and dropping below 
detection at the edge of the map. 
The emission of the other lines is rather compact. 
The shape of the PACS spectrometer beam is not known well 
enough to measure an accurate source size if the source were slightly extended. 
Hence we cannot recover spatial information. 
To derive total line intensities, we integrated the fluxes over a circular aperture 
of diameter 30$^{\prime\prime}$ to encompass all the emission. 
The line intensities are listed in Table~\ref{table:lines}.
Observations of the [O~{\sc iii}], [C~{\sc ii}], and [O~{\sc i}]~63 and 
145~$\mu$m lines were also performed with the LWS instrument 
onboard {\it ISO} \citep{bergvall-2000,brauher-2008,vasta-2010}. 
The PACS intensities of the [O~{\sc i}]~63~$\mu$m and [O~{\sc iii}] lines 
agree within 40\% with the LWS observations, although they are 
found to be lower with PACS. This may be attributed to a beam effect, 
the LWS beam size being $\sim$80$^{\prime\prime}$. 
However, we find that the [C~{\sc ii}] intensity is two times brighter with PACS compared to LWS. 
To understand this discrepancy, we have reduced the PACS data with different flat-field 
corrections, as well as with background normalization (PACS Data Reduction Guide). 
All these methods give the same final [C~{\sc ii}] flux as quoted in Table~\ref{table:lines} 
within 10\%, ruling out a mis-calibration from PACS. 
Therefore we attribute the difference between the PACS 
and LWS fluxes to a calibration problem from the LWS instrument.

\begin{center}
\begin{table*}[ht]
  \caption{{\it Spitzer} and {\it Herschel} line fluxes.}
  \hfill{}
  \begin{tabular}{l c c c c c c}
    \hline\hline
    \multicolumn{1}{l}{Line} & 
    \multicolumn{1}{c}{Wavelength} &
    \multicolumn{1}{c}{Flux$^{(a)}$} &
    \multicolumn{1}{c}{Uncertainty} &
    \multicolumn{1}{c}{Excit. potential$^{(b)}$} &
    \multicolumn{1}{c}{Excit. temperature$^{(c)}$} &
    \multicolumn{1}{c}{Critical density$^{(d)}$} \\
    \multicolumn{1}{l}{ } & 
    \multicolumn{1}{c}{($\mu$m)} &
    \multicolumn{1}{c}{(10$^{-16}$ W~m$^{-2}$)} &
    \multicolumn{1}{c}{(10$^{-16}$ W~m$^{-2}$)} &
    \multicolumn{1}{c}{(eV)} &
    \multicolumn{1}{c}{(K)} &
    \multicolumn{1}{c}{(cm$^{-3}$)} \\
    \hline
    	{\it Spitzer}/IRS 		& 	& 	& 	& 	& 	& 	 \\
    \hline
	$\rm{[}$S~{\sc iv}$\rm{]}$		& 10.51	& 5.37	& 0.66	& 34.79	& 1 369	& 5 $\times \rm{10^{4}}$ [e] \\
	$\rm{[}$Ne~{\sc ii}$\rm{]}$		& 12.81	& 3.53	& 0.43	& 21.56	& 1 123	& 7 $\times \rm{10^{5}}$ [e] \\
	$\rm{[}$Ne~{\sc iii}$\rm{]}$	& 15.56	& 12.3	& 1.5		& 40.96	& 925	& 3 $\times \rm{10^{5}}$ [e] \\
	$\rm{[}$S~{\sc iii}$\rm{]}$		& 18.71	& 5.79	& 0.72	& 23.34	& 769	& 2 $\times \rm{10^{4}}$ [e] \\
	$\rm{[}$S~{\sc iii}$\rm{]}$		& 33.48	& 5.85	& 0.69	& ''		& 430	& 7 $\times \rm{10^{3}}$ [e] \\
	$\rm{[}$Si~{\sc ii}$\rm{]}$		& 34.82	& 6.21	& 0.75	& 8.15	& 413	& 3 $\times \rm{10^{5}}$ [H], 1 $\times \rm{10^{3}}$ [e] \\
	Humphreys~$\alpha$		& 12.37	& 0.208	& 0.045	& 13.60	& 1 163	& - \\
	H{\scriptsize 2}	(0,0) S(0)		& 28.22	& $<$ 0.62	& -	& 4.48	& 510	& 7 $\times \rm{10^{2}}$ [H] \\ 
	H{\scriptsize 2}	(0,0) S(1)		& 17.03	& 0.168	& 0.055	& ''		& 1015	& 2 $\times \rm{10^{4}}$ [H] \\ 
	H{\scriptsize 2} (0,0) S(2)		& 12.28	& 0.101	& 0.013	& '' 		& 1682	& 2 $\times \rm{10^{5}}$ [H] \\
	H{\scriptsize 2}	(0,0) S(3)		& 9.66	& 0.121	& 0.018	& '' 		& 2504	& 9 $\times \rm{10^{5}}$ [H] \\ 
	$\rm{[}$Fe~{\sc iii}$\rm{]}$		& 22.93	& 0.54	& 0.20	& 16.19	& 627	& 1 $\times \rm{10^{5}}$ [e] \\
	$\rm{[}$O~{\sc iv}$\rm{]}$		& 25.89	& 0.49	& 0.18	& 54.94	& 555	& 1 $\times \rm{10^{4}}$ [e] \\
	$\rm{[}$Fe~{\sc ii}$\rm{]}$		& 25.99	& $<$ 0.25 	& -	& 7.90	& 554	& 2 $\times \rm{10^{6}}$ [H], 1 $\times \rm{10^{4}}$ [e] \\
	$\rm{[}$Ar~{\sc iii}$\rm{]}$		& 8.99	& 0.99	& 0.33	& 27.63	& 2 060	& 3 $\times \rm{10^{5}}$ [e] \\
	$\rm{[}$Ar~{\sc ii}$\rm{]}$		& 6.99	& 0.60	& 0.15	& 15.76	& 1 600	& 4 $\times \rm{10^{5}}$ [e] \\
	$\rm{[}$Ne~{\sc v}$\rm{]}$		& 14.32	& $<$ 0.080	& -	& 97.12	& 592	& 3 $\times \rm{10^{4}}$ [e] \\
    \hline
    	{\it Herschel}/PACS	 	&	&	&	&	& \\
    \hline
	$\rm{[}$C~{\sc ii}$\rm{]}$		& 157.74	& 7.4		& 2.2		& 11.26	& 91		& 3 $\times \rm{10^{3}}$ [H], 50 [e] \\
	$\rm{[}$O~{\sc iii}$\rm{]}$		& 88.36	& 18.0	& 5.4		& 35.12	& 163	& 5 $\times \rm{10^{2}}$ [e] \\
	$\rm{[}$O~{\sc i}$\rm{]}$		& 63.18	& 6.7		& 2.0		& - 		& 228	& 5 $\times \rm{10^{5}}$ [H] \\
	$\rm{[}$O~{\sc i}$\rm{]}$		& 145.52	& 0.56	& 0.17	& - 		& 99		& 1 $\times \rm{10^{5}}$ [H] \\
	$\rm{[}$N~{\sc iii}$\rm{]}$		& 57.32	& 2.50	& 0.78	& 29.60	& 251	& 3 $\times \rm{10^{3}}$ [e] \\
	$\rm{[}$N~{\sc ii}$\rm{]}$		& 121.9	& 0.31	& 0.10	& 14.53	& 118	& 3 $\times \rm{10^{2}}$ [e] \\
	$\rm{[}$N~{\sc ii}$\rm{]}$		& 205.18	& $<$ 0.29	& -	& '' 		& 70		& 45 [e] \\
    \hline\hline
  \end{tabular}
  \hfill{}
  \newline
  The {\it ISO}/LWS fluxes can be found in \cite{bergvall-2000}, \cite{brauher-2008}, and \cite{vasta-2010}. \\
    ${(a)}$ Upper limits are given as the 3-$\sigma$ uncertainty from the line fit. \\
    ${(b)}$ Excitation potential. Energy to create an ion. For H$\rm{_2}$, it is the energy to photodissociate the molecule. \\
    ${(c)}$ Excitation temperature. Temperature T=$\Delta$E/k required to populate the transition level. \\
    ${(d)}$ Critical density n$_{crit}$ noted [H] for collisions with hydrogen atoms (T=100~K and T=300~K for the H$\rm{_2}$ lines), 
    [e] with electrons (T=10~000~K). Values are from \cite{malhotra-2001}, \cite{giveon-2002}, and \cite{kaufman-2006}.
  \label{table:lines}
\end{table*}
\end{center}

\subsection{Spectral line description}
The IRS and PACS spectra probe the ionised and neutral media.
The IRS spectrum shows a wealth of MIR fine-structure lines, the brightest being 
(in order) [Ne~{\sc iii}]~15.6~$\mu$m, [Si~{\sc ii}]~34.8~$\mu$m, [S~{\sc iii}]~18.7 
and 33.5~$\mu$m, [S~{\sc iv}]~10.5~$\mu$m, and [Ne~{\sc ii}]~12.8~$\mu$m.
PAH features at 6.2, 7.7, 8.6, 11.2, and 12.8~$\mu$m are also present. 
The PACS spectrum also shows bright FIR lines, especially [O~{\sc iii}], [C~{\sc ii}], 
[O~{\sc i}] 63~$\mu$m, and [N~{\sc iii}]. 
The [O~{\sc iii}] line is the brightest of all, with [O~{\sc iii}]/[C~{\sc ii}]~=~2.4. 
While [C~{\sc ii}] is the brighest FIR fine-structure line in normal and dusty starburst 
galaxies \citep{brauher-2008}, we find that this is often not the case in low-metallicity 
dwarf galaxies \citep{hunter-2001,cormier-2010,cormier-2012b,madden-2012b}.

The young star-forming nature of this galaxy also agrees with intense [Ne~{\sc iii}] (41.0~eV), 
[O~{\sc iii}] (35.1~eV), [S~{\sc iv}] (34.8~eV), and [N~{\sc iii}] (29.6~eV) lines observed, 
arising from relatively highly ionised regions. [O~{\sc iv}] (54.9~eV) at 25.9~$\mu$m, 
which requires a relatively hard radiation field (Sect.~\ref{sect:xrays}), is also detected. 
The [Ne~{\sc iii}]/[Ne~{\sc ii}] ratio is 3.5, [S~{\sc iv}]/[S~{\sc iii}] is 0.9, and 
[N~{\sc iii}]/[N~{\sc ii}] is 8. These ratios are indicative of a hard radiation field 
typical of BCDs \citep{thornley-2000,madden-2006}. 
While [C~{\sc ii}] can come from both the diffuse ionised phase and the neutral phase, 
[O~{\sc i}] traces only the neutral phase. 
The [O~{\sc i}]~63~$\mu$m/[C~{\sc ii}] ratio is 0.9, similar to 
that found in other dwarf galaxies and dustier galaxies 
\citep{hunter-2001,brauher-2008}. 
Table~\ref{table:lines} lists the ionisation energies of the observed lines, 
as well as the excitation temperatures and critical densities (n$_{crit}$) 
of the transitions with collisions with H and/or e$^-$.

Comparing line luminosities to the TIR luminosity, we find 
L$\rm{_{[C~II]}}$/L$\rm{_{TIR}}$~=~0.1\%, 
L$\rm{_{[O~III]}}$/L$\rm{_{TIR}}$~=~0.3\%, and 
L$\rm{_{[O~I]}}$/L$\rm{_{TIR}}$~=~0.1\%. 
Altogether, the FIR lines represent 0.6\% of the TIR luminosity; and 
when adding the MIR lines, they represent 1.2\% of L$\rm{_{TIR}}$. 
Together these lines are responsible for most of 
the gas cooling of the ISM. 
Such line-to-TIR ratios are typical of star-forming galaxies 
\citep{malhotra-2001, brauher-2008}\footnote{
Note that they use $\rm{L_{FIR}}$, determined from {\it IRAS} fluxes with the \cite{helou-1988} formula, 
and that $\rm{L_{TIR}}$ estimated as $\rm{2 \times L_{FIR}}$ \citep{hunter-2001} can be different 
from the true $\rm{L_{TIR}}$ when including submm observations. 
}. 
Similar L$\rm{_{[C~II]}}$/L$\rm{_{TIR}}$ ratios are found in 
several high-redshift galaxies \citep{maiolino-2009,stacey-2010}. 
\cite{croxall-2012} find $\rm{L_{[C~II]}/L_{TIR}\sim0.5\%}$ and 
$\rm{L_{[O~I]}/L_{TIR}\sim0.1\%}$ in spatially resolved regions 
of the nearby spiral galaxy NGC~4559. 
There is observational evidence that L$\rm{_{[C~II]}}$/L$\rm{_{TIR}}$ 
tends to decrease in galaxies with warmer FIR-color, 
f$_{\nu}$(60~$\mu$m)/f$_{\nu}$(100~$\mu$m). Among other effects, 
this may be attributed to a less efficient photoelectric heating due to 
charged grains \citep{malhotra-2001,croxall-2012}, or to a UV-photon 
screening from large amounts of dust as in ultra-luminous IR galaxies (ULIRGs) 
\citep{abel-2009,gracia-2011}. With observed FIR-color $\sim$1, 
Haro\,11 falls within the large spread of L$\rm{_{[C~II]}}$/L$\rm{_{TIR}}$ 
values and is not particularly [C~{\sc ii}] deficient nor [C~{\sc ii}] excessive, 
as often seen in low metallicity galaxies \citep{madden-2012b}.

\subsection{Photometry data}
For the discussion on the SED of Haro\,11 in Section~\ref{sect:sed}, 
we use broad-band data available from the X-ray to the radio regime, 
to assess how the models perform in reproducing the continuum emission. 
The photometry is taken from NED. 
In addition, we use the {\it Spitzer}/MIPS photometry 
from \cite{bendo-2012}, the {\it Herschel}/PACS and SPIRE photometry 
from \cite{remy-2012}, and the {\it Spitzer}/IRAC and LABOCA 
870~$\mu$m photometry from \cite{galametz-2009}.
All fluxes are summarized in Table~\ref{table:phot}.

\begin{center}
\begin{table}[ht]
  \caption{Photometry data.}
  \hfill{}
  \begin{tabular}{l c c c c}
    \hline\hline
    \multicolumn{1}{l}{Instrument} & 
    \multicolumn{1}{c}{Wavelength} &
    \multicolumn{1}{c}{Flux} &
    \multicolumn{1}{c}{Uncertainty} &
    \multicolumn{1}{c}{Reference} \\
    \multicolumn{1}{l}{ } & 
    \multicolumn{1}{c}{} &
    \multicolumn{1}{c}{(Jy)} &
    \multicolumn{1}{c}{(Jy)} &
    \multicolumn{1}{c}{} \\
    \hline
	Chandra		& 2-10~keV	& 9.17e-09	& 1.03e-09	&  (1) \\
	Chandra		& 0.5-2~keV	& 3.21e-08	& 0.43e-08 	&  (1) \\
	ROSAT		& 0.1-2~keV	& 4.41e-08	& 0.55e-08	&  (2) \\
	FUSE		& 1150~\AA	& 2.30e-03	& 0.33e-03	&  (3) \\
	GALEX		& 1530~\AA	& 3.63e-03	& 7e-05	&  (4) \\
	GALEX		& 2315~\AA	& 4.37e-03	& 4e-05	&  (4) \\
	Cousins		& 0.439~$\mu$m 	& 8.03e-03	& 0.69e-03	&  (5) \\
	Cousins		& 0.639~$\mu$m	& 0.012	 	& 0.001	&  (5) \\
	2MASS		& 1.25~$\mu$m	& 0.013 		& 0.001	&  (6) \\
	2MASS		& 1.64~$\mu$m	& 0.013 		& 0.001	&  (6) \\
	2MASS		& 2.17~$\mu$m	& 0.014	 	& 0.001	&  (6) \\
	IRAC		& 3.6~$\mu$m		& 0.023	& 0.002	&  (7) \\
	IRAC		& 4.5~$\mu$m		& 0.029	& 0.003	&  (7) \\
	IRAC		& 5.8	~$\mu$m		& 0.073	& 0.007	&  (7) \\
	IRAC		& 8~$\mu$m		& 0.177	& 0.018	&  (7) \\
	IRAS			& 12~$\mu$m		& 0.52	& 0.02	&  (8) \\
	MIPS		& 24~$\mu$m		& 2.51	& 0.03	&  (9) \\
	IRAS			& 25~$\mu$m		& 2.49	& 0.03	&  (8) \\
	IRAS			& 60~$\mu$m		& 6.88	& 0.04	&  (8) \\
	MIPS		& 70~$\mu$m		& 4.91	& 0.49	&  (9) \\
	PACS		& 70~$\mu$m		& 6.08	& 0.18	&  (10) \\
	IRAS			& 100~$\mu$m		& 5.04	& 0.03	&  (8) \\
	PACS		& 100~$\mu$m		& 4.97	& 0.15	&  (10) \\
	MIPS		& 160~$\mu$m		& 2.01	& 0.24	&  (9) \\
	PACS		& 160~$\mu$m		& 2.43	& 0.12	&  (10) \\
	SPIRE		& 250~$\mu$m		& 0.633	& 0.042	&  (10) \\
	SPIRE		& 350~$\mu$m		& 0.231	& 0.013	&  (10) \\
	SPIRE		& 500~$\mu$m		& 0.092	& 0.008	&  (10) \\
	LABOCA		& 870~$\mu$m		& 0.040	& 0.006	& (7) \\
    \hline\hline
  \end{tabular}
  \newline
  \hfill{}
(1) \cite{hayes-2007}; (2) \cite{tajer-2005}; (3) \cite{grimes-2009}; 
(4) \cite{iglesias-2006}; (5) \cite{lauberts-1989}; 
(6) \cite{jarrett-2000}; (7) \cite{galametz-2009}; (8) \cite{soifer-2003}; 
(9) \cite{bendo-2012}; (10) \cite{remy-2012}.
  \newline
  \label{table:phot}
\end{table}
\end{center}

\section{Modeling strategy}
\label{sect:model}

\subsection{Identification of the model components}
We aim to model the IR emission from all 17 detected lines 
(except the MIR H$\rm{_2}$ lines which will be modeled in a coming paper) 
listed in Table~\ref{table:lines} to understand the conditions of the various media and 
the dominant processes at work in this low-metallicity starburst. 
The ISM is very heterogeneous, and over the scale of a galaxy, 
various phases with a wide range of conditions are mixed together.

In our spectral dataset, we notice that Haro\,11 is typified by very intense 
[Ne~{\sc iii}], [Ne~{\sc ii}], [S~{\sc iii}], [S~{\sc iv}], [O~{\sc iii}], and [N~{\sc iii}] lines. 
These high-excitation lines require a hard radiation field which is 
naturally encountered in a compact H~{\sc ii} region. 
\revised{
Stars in compact H~{\sc ii} regions are young, hence their UV radiation is harder,
and SSCs are likely to contain massive stars which contribute to the production 
of hard photons. 
}
Moreover, the high critical densities of these lines (except the 
[O~{\sc iii}] line) favor their origin at the same location. 
Therefore, the intensity of these lines demonstrates the importance 
of the dense H~{\sc ii} region in the global spectrum of Haro\,11.
Moreover, the {\it Herschel} observations show bright [O~{\sc i}] and 
[C~{\sc ii}] lines, which are usual tracers of the neutral gas. 
In particular, the intensity of the [O~{\sc i}] lines indicates the 
presence of dense PDRs. 
These two dense (ionised and neutral) phases will be investigated 
first since they account for the emission of most of the spectral lines. 
Derived physical conditions are presented in Section~\ref{sect:results}. 

With low critical density ($\sim$300~cm$^{-3}$) and low ionisation 
potential (14.5~eV), the [N~{\sc ii}]~122~$\mu$m line is a tracer 
of low-excitation diffuse gas rather than the dense H~{\sc ii} region. 
The intensity of [N~{\sc ii}] demonstrates the need for an additional 
diffuse low-ionisation component that is investigated in Section~\ref{sect:diffus}. 
Our subsequent modeling also shows that this diffuse phase is 
the main contributor to the [Ne~{\sc ii}] and [C~{\sc ii}] lines. 

Therefore, the three main phases that we identify as necessary 
model components to account for the observed emission lines are: 
1)~a dense ionised phase (Sect.~\ref{sect:compact}),
2)~PDR (Sect.~\ref{sect:pdr}), 
3)~a diffuse low-ionisation medium (Sect.~\ref{sect:diffus}). 

In addition, we discuss the possible presence of other model components 
that may influence the line predictions. 
The detection of the high-excitation [O~{\sc iv}] line is the driver 
of a discussion on the role of X-rays and XDRs in Section~\ref{sect:xrays}. 
The [O~{\sc iii}] line, which is the brighest of all MIR-FIR lines, 
has high excitation potential (35.1~eV) and low critical density 
($\sim$500~cm$^{-3}$). Although it is accounted for by the dense 
H~{\sc ii} model, it may originate as well from diffuse high-excitation medium 
that we do not model but briefly discuss in Section~\ref{sect:oiii}. 
Finally, we investigate the influence of magnetic fields on the cloud 
density profile and their impact on the PDR predictions in 
Section~\ref{sect:bmag}.

\subsection{Method}
We use the 1D spectral synthesis code Cloudy v10.00, last described 
by \cite{ferland-1998}, which includes photoionisation and photodissociation 
treatment \citep{abel-2005}, to model the multiple phases from which 
the emission lines originate. 
Since Haro\,11 is not resolved in most of the {\it Spitzer} and {\it Herschel} bands, 
we have no spatial constraints. 
Thus, the model is reduced to a central source of energy surrounded by a spherical cloud. 
In the case of Haro\,11, this central source is a starburst. 
The dense ionised gas that is closest to the central source with respect to the 
other phases that we model separately will yield the best 
constraints on the properties of the stellar cluster that is powering the model. 
Therefore we first derive the best fit model concentrating on the dense 
H~{\sc ii} region diagnostics, which are the [Ne~{\sc ii}]~12.8~$\mu$m, 
[Ne~{\sc iii}]~15.6~$\mu$m, [S~{\sc iv}]~10.5~$\mu$m, [S~{\sc iii}]~18.7~$\mu$m 
and 33.5~$\mu$m, and the [N~{\sc iii}]~57~$\mu$m lines. 
Moreover, the interpretation of these ionic lines 
is less ambiguous than, for example, the [C~{\sc ii}] line. 
Following this step, we will assess what this model predicts for the PDR phase.

The starburst ionises the inner edge of the cloud, where the H~{\sc ii} region begins, 
and the radiative transfer is computed step by step progressively into the cloud. 
The distance from the source to the inner edge of the cloud is the inner radius (r$\rm{_{in}}$). 
Depending on where the calculation is stopped, the ionised, atomic, and molecular phases 
can be treated as the model is computed within one Cloudy model. 
Here we stop the models until the molecular fraction H$\rm{_{2}}$/H reaches 50\% 
such that both the ionised and PDR phases are computed within the same model.
We opt to impose pressure equilibrium throughout the models. 
This is particularly important for the density profile between different phases.
We will come back to this assumption in Sect.~\ref{sect:pdr}.

\subsubsection{Parameter grid of the dense H~{\sc ii} region}
\begin{figure}[!htp]
\centering
\includegraphics[clip,width=8.8cm]{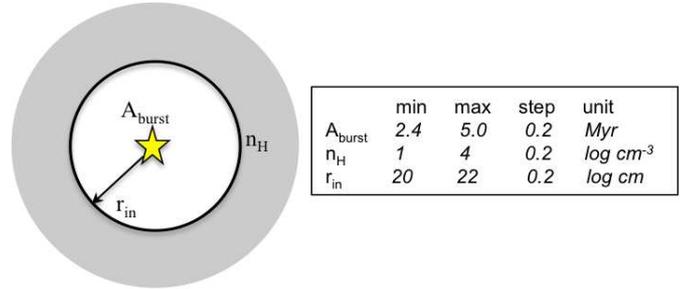}
\caption{
Geometry of the Cloudy model with 
the three free parameter that we vary in the grid:
the age of the burst (A$\rm{_{burst}}$), inner radius 
(r$\rm{_{in}}$), and initial hydrogen density (n$\rm{_{H}}$). 
Parameter boundaries and step sizes are also indicated. 
}
\label{fig:cloud}
\end{figure}

We vary the following 3 physical parameters in order to calculate our model grid:
the age of the burst (A$\rm{_{burst}}$), the initial hydrogen density at the 
inner edge of the cloud (n$\rm{_H}$), and r$\rm{_{in}}$, where the calculation starts 
(see Fig.\ref{fig:cloud} for an illustration). 
The age is varied from 2.4 to 5.0~Myr (Sect.~\ref{sect:age}), 
r$\rm{_{in}}$ from $10^{20.0}$ to $10^{22.0}$~cm (or 30~pc to 3~kpc, Sect.~\ref{sect:rin}), 
and n$\rm{_H}$ from $10^{1}$ to $10^{4}$~cm$^{-3}$ (Sect.~\ref{sect:hden}). 
The step size for A$\rm{_{burst}}$ is 0.2~Myr (linear step), 
and for r$\rm{_{in}}$ and n$\rm{_H}$, we use 0.2 step sizes on a logarithmic scale.

\subsubsection{$\chi^2$ evaluation}
In order to find a solution for this ionised phase, we aim first to reproduce 
as best as possible the intensities of the ionic lines. 
For each individual model, the goodness of the fit is evaluated by computing 
the reduced $\chi^2$. The $\chi^2$ is measured from a set of selected lines.
The lines we consider to constrain the fit are the following: 
[Ne~{\sc ii}]~12.8~$\mu$m, [Ne~{\sc iii}]~15.6~$\mu$m, 
[S~{\sc iv}]~10.5~$\mu$m, [S~{\sc iii}]~18.7~$\mu$m, and 33.5~$\mu$m, 
because they are the brightest and relatively well understood, 
as well as the [N~{\sc iii}]~57~$\mu$m line from PACS 
which is also bright and expected to arise from the same medium. 
With 6 observations and 3 free parameters to fit, 
the degree of freedom is $\rm{N = n_{obs} - n_{param} = 3}$.
Best fit models are determined by minimising the reduced $\chi^2$. 
There are several lines that originate in part from the dense ionised gas 
that we do not consider when calculating the $\chi^2$, although we do 
discuss how the model predictions of these lines compare to the observations. 
We do not consider the high-excitation [O~{\sc iv}] line because it is faint 
and can be excited by several mechanisms (see Sect.~\ref{sect:xrays}).
We do not include the [O~{\sc iii}] and [N~{\sc ii}] lines here as constraints for the dense 
H~{\sc ii} region because our subsequent modeling shows that a fraction 
of these lines arises from a low density ionised medium which 
we consider in a later step (see Sect.~\ref{sect:diffus}).

\subsection{Setting the input fixed parameters}
\subsubsection{Incident radiation field}
\label{sect:irf}
We use Starburst99 v6.0.2 \citep{leitherer-2010} to reproduce the 
incident, wavelength-dependent radiation field density 
of Haro\,11. Following the UV-optical studies by \cite{hayes-2007} and \cite{grimes-2007},
we assume a Salpeter IMF from 0.1 to 100~M$_{\odot}$ and 
stellar tracks from Padova AGB with metallicity $\rm{Z = 0.004}$.
According to \cite{adamo-2010}, the stellar population of Haro\,11 is very young. 
The present starburst started less than 40~Myr ago, and shows a formation peak 
at about 3.5~Myr. 
Knot B is dominated by a young starburst of 3.5~Myr, and knot C contains 
an older starburst of 9.5~Myr (Fig.~\ref{fig:acs}).
Therefore, we select the stellar spectrum of an instantaneous burst of a few Myr, 
letting the age of the burst be a free parameter, varying from 2.4 to 5.0~Myr. 
We also add to this the spectrum of an instantaneous burst of 9.5 Myr to 
account for the slightly older stellar population. We keep this age fixed since 
the ionising part of the stellar spectrum does not vary significantly for ages larger 
than 5~Myr, and will therefore not significantly affect the model predictions of the ionised gas.
This starburst spectrum dictates the shape of the input energy source in Cloudy. 
The incident radiation field on the cloud is shown in purple in Figure~\ref{fig:sed}.
For the input luminosity of the central source, we take the total observed infrared 
luminosity of Haro\,11: $\rm{L_{TIR} = 1.4 \times 10^{11}~L_{\odot}}$. 
This approximation assumes an energy balance between the UV and IR domains, 
i.e. that all of the UV photons are reprocessed in the IR by the gas and the dust.
The emergent SED (bottom panel of Fig.~\ref{fig:sed}) shows that 
this approximation is reasonable since only 20\% of the total emergent 
luminosity is emitted in the UV-optical, and 80\% in the IR. 
We discuss in more details the energy balance in Sect~\ref{sect:sed}.

Haro\,11 is detected in X-rays \citep[e.g.][]{grimes-2007}. 
Previous studies on X-ray observations indicate the presence of soft 
($\rm{L_{0.5-2 kev} = 1.9 \times 10^{8}~L_{\odot}}$) and 
hard ($\rm{L_{2-10 kev} = 2.6 \times 10^{8}~L_{\odot}}$) X-rays. 
Therefore we add to the input SED an X-ray component that reproduces 
the shape of the observed X-ray spectrum and of total luminosity 
$\rm{5 \times 10^{8}~L_{\odot}}$.
The effects of the X-rays in the models are discussed in Section~\ref{sect:xrays}.

\subsubsection{Elemental abundances}
The elemental abundances directly impact on the predicted line intensities. 
To match the observed abundances in Haro\,11 as closely as possible, 
we use the abundances in the literature. 
Abundances of oxygen, nitrogen, neon, sulphur, and argon were obtained 
in knots B and C of Haro\,11 by \cite{guseva-2012} using {\it VLT}/X-shooter 
observations in the optical. 
In particular, they find an oxygen abundance 12+log(O/H) of 8.1 in knot~B 
and 8.3 in knot~C, giving a O/H twice as large as the value 12+log(O/H)=7.9 
previously found by \cite{bergvall-2002}, 
altering the determination of the metallicity from $\sim$1/6 to 1/3~Z$_{\odot}$. 
We use the abundances in \cite{guseva-2012}, averaged over the two knots, 
which have similar luminosities, for our calculations. 
Uncertainties on these abundances are 10\%. 
For neon and sulphur, \cite{guseva-2012} find similar values to those found 
in \cite{wu-2008a}, derived from {\it Spitzer} data.
The relatively high neon abundance is typical of BCDs \citep{izotov-1999}.
For nitrogen, such high abundance is observed in a few BCDs 
\citep[e.g.][]{thuan-1996,izotov-1998,pustilnik-2004},
and can be explained by the presence of Wolf-Rayet stars \citep{bergvall-2002}, 
which inject matter enriched in N and C into the ISM. 
For the other elements, we use gas-phase relative abundances typical of H~{\sc ii} regions, 
based on an average of the abundances in the Orion Nebula determined by 
\cite{baldwin-1991,rubin-1991,osterbrock-1992,rubin-1993}, and scaling them 
to the metallicity of Haro\,11. The resulting carbon abundance is 
$\mathrm{log(C/O)=-0.15}$, which is relatively high compared to \cite{izotov-1999}, 
where $\mathrm{log(C/O)\sim-0.5}$. However, \cite{bergvall-2006} find 
that the ratio of C and O column densities is $\sim-0.3$ (log) 
from the analysis of absorption lines in IUE and FUSE data. 
Such high value may indicate an enrichment in C and agrees with 
our adopted carbon abundance (assuming Orion-like $\mathrm{C/O}$) 
within the uncertainties. For silicon, its abundance $\mathrm{Si/O}$ is low 
in the Orion abundance set compared to that usually found in BCDs, 
therefore we use the average abundance of $\mathrm{log(Si/O)=-1.5}$ 
from \cite{izotov-1999}. 
We consider the uncertainty on these theoretical abundances to be of a factor of 2. 
Table~\ref{table:ab} summarizes the elemental abundances (and uncertainties) 
we use in Cloudy. The lack of measured abundances for the other elements 
limits their use to constrain the models (e.g. Fe, Si).

\begin{center}
\begin{table}[!htp]
  \caption{Elemental abundances set in Cloudy.}
  \begin{tabular}{l l}
    \hline\hline
    \multicolumn{1}{l}{Gas-phase composition} & 
    \multicolumn{1}{l}{Abundance (uncertainty) in log} \\
    \hline
	O/H	&	-3.79	 (0.14)	\\
	N/H	&	-4.64	 (0.12)	\\
	Ne/H&	-4.44	 (0.07)	\\
	S/H	&	-5.57	 (0.10)	\\
	Ar/H	&	-6.20	 (0.13)	\\
	C/H	&	-3.92	 (0.3) 	\\
	Si/H	&	-5.80	 (0.3) 	\\
	Fe/H	&	-5.92 (0.3)		\\
    \hline
    \multicolumn{1}{l}{Grain composition}  &
    \multicolumn{1}{l}{Mass fraction}    \\
    \hline
	Carbonaceous &	36\%		\\
	Silicate &	60\%		\\
	PAH &	4\%		\\
    \hline
	Dust-to-Gas mass ratio &	$\rm{2.64 \times 10^{-3}}$		\\
    \hline \hline
  \end{tabular}
  \begin{flushleft}
	Abundances of oxygen, nitrogen, neon, sulphur, and argon are from \cite{guseva-2012}, 
	silicon from \cite{izotov-1999}, and the others are average abundances of the 
	Orion Nebula determined by \cite{baldwin-1991,rubin-1991,osterbrock-1992,rubin-1993}, 
	scaled to the metallicity of Haro\,11.\\
	The main elemental composition O:C:Si:Fe in the grains is 4:9:1:1.
  \end{flushleft}
  \label{table:ab}
\end{table}
\end{center}

\subsubsection{Dust properties}
\label{sect:dustprop}
It is important to accurately set the dust properties for two reasons: 
(1) dust plays an important role in heating the gas through the photoelectric 
effect, which depends on the dust abundance and grain size distribution, and 
(2) the radiative transfer is strongly influenced by the properties of the grains. 
We first tried using the standard MRN grain size distribution \citep*{mathis-1977}, 
which failed to reproduce the observed dust emission in the MIR, 
falling below the photometry data by more than an order of magnitude. 
The MRN distribution underpredicts the number of small grains 
(minimum grain radius of 5~nm) 
and is usually not appropriate for modeling the MIR emission in galaxies. 
Standard dust models such as \cite{draine-2001} and \cite{zubko-2004} 
do include more small grains than the standard MRN model. 
Dust modeling of dwarf galaxies \citep[e.g.][]{galliano-2003,galliano-2005,galametz-2011} 
also demonstrates the need for an increased abundance of smaller grains.
Therefore, in Cloudy, we use the MRN distribution 
extended down to 1.1~nm to account for the very small grains. 
\revised{
Other grain size distributions have been used to model the IR emission of galaxies, 
such as grain shattering distributions \citep{dopita-2005a}, that we discuss in 
Section~\ref{sect:sedmir}. 
}
We also use PAHs as described in \cite{abel-2008}.
Both grains and PAH abundances are scaled to the metallicity of Haro\,11 
(1/3~Z$_{\odot}$), fixing the dust-to-gas mass ratio (D/G) to 
$\rm{2.64 \times 10^{-3}}$ ($\sim$1/400, the Galactic D/G is $\sim$1/150).
This yields a ratio of visual extinction to hydrogen column density, 
A$_V$/N(H), of $\rm{10^{-22}~mag~cm^{-2}}$.

\subsection{Limitations of the approach}
\label{sect:limits}
Our first limitation lies on the geometry of our modeling that we aim to 
constrain as much as possible with diagnostics from a large set of spectral lines. 
Some of the derived parameters of the compact H~{\sc ii} region are
strongly dependent on the (simple) geometry we impose on the model,
which we will see is a recurring theme throughout the paper. 
The parameters that are best constrained, irrespective of the geometry,
are the electron density, temperature and mass of ionised gas. 
As found in Section~\ref{sect:rin}, the geometrically thin ionised gas shell, 
in combination with the large inner radius ($\sim$1~kpc) that describes the 
model solution of the dense H~{\sc ii} region, seems physically implausible. 
Despite the geometry of the model, we refer to it with the term compact 
since this model component represents what is observationally 
described as a compact H~{\sc ii} region. This is a direct result
of assuming a single region with a single central source. 
Although we are aware that the knots~B and C of Haro\,11 are 
spatially separated in the optical, we do not have the spatial resolution 
in the {\it Spitzer} and {\it Herschel} observations to treat them separately. 
If, for example, we were to construct a model as the sum of several regions --
as seems reasonable when perusing the optical images -- the inner
radius of each region would have to decrease in order to obtain a similar 
effective $U$ parameter, while the shell thickness would increase in order 
to obtain the same gas mass, thus evolving towards a model with a more
geometrically thick layout. 
It is impossible to obtain reliable and reproducible results using such a
complex model, without better observational constraints on the actual
source distribution, i.e. spatially resolved spectroscopy. In this
paper we focus on the derived parameters that are least sensitive on
the actual assumed geometry. 
In Section~\ref{sect:sed}, we discuss the spectral energy distribution 
resulting from the combination of models of the various phases in terms 
of energy balance, but we do not pretend to accurately fit the total SED 
because of these geometry issues.

Our second limitation lies in the evalutation of the goodness of the fits. 
It is mainly weighted by the elemental abundance uncertainties, 
which can be up to a factor of 2, and by the observational 
uncertainties (up to 40\%). Therefore, when comparing the model predictions 
to observed intensities, we aim for a match within a factor of 2.

\section{A model of the H~{\sc ii} region and PDR: $\mathscr{C}$}
\label{sect:results}
\subsection{The compact H~{\sc ii} region}
\label{sect:compact}
\begin{figure}[!tph]
\centering
\includegraphics[width=8.8cm]{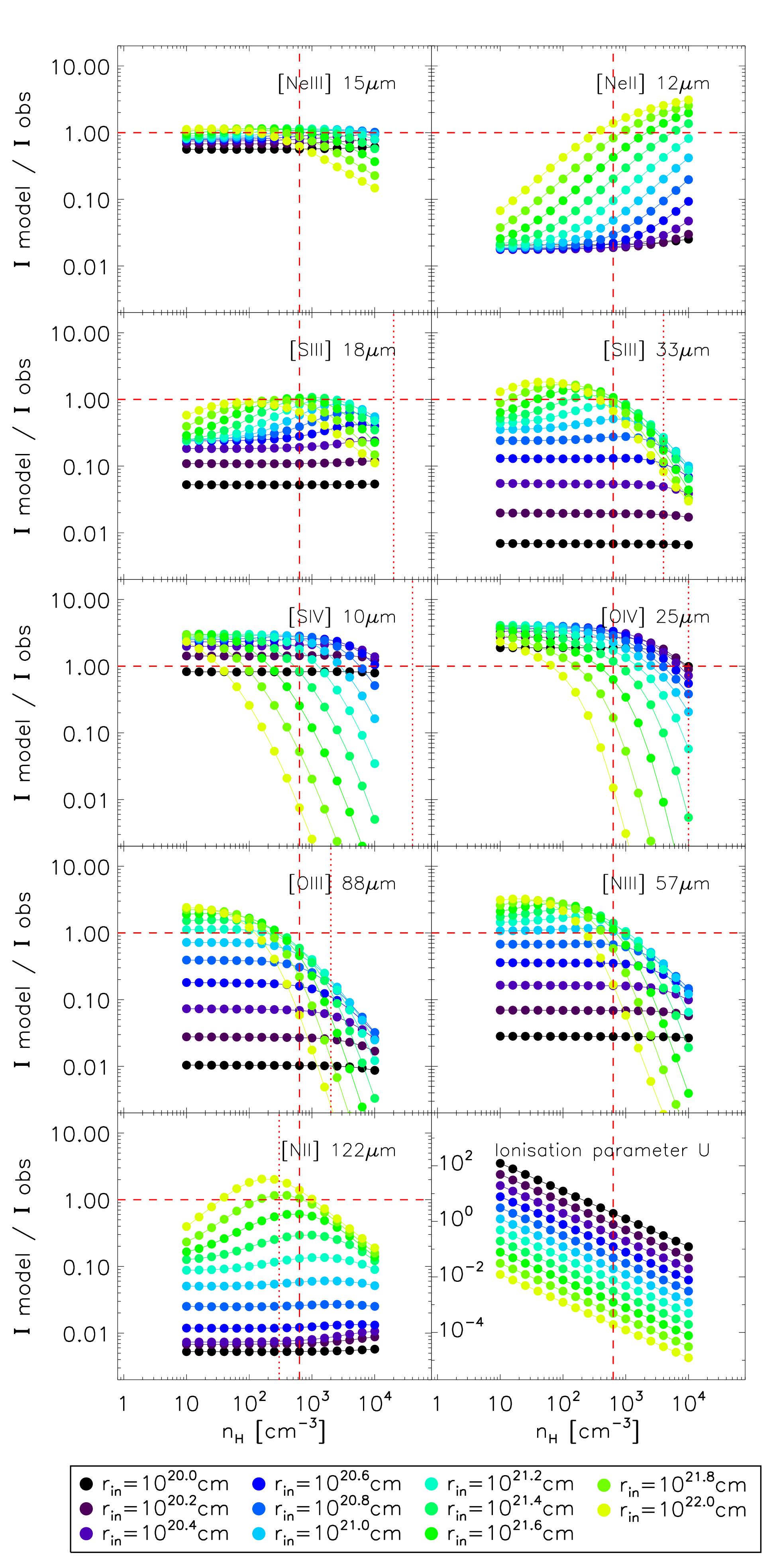}
\caption{Predicted intensities for model $\mathscr{C}$ (I$_{model}$) 
normalised by their observed value (I$_{obs}$) for all ionic lines considered, 
as a function of n$\rm{_{H}}$ and r$_{\rm in}$ for a given starburst age (of 3.7~Myr). 
The vertical dashed line shows the best fit model density 
n$\rm{_H}=$10$^{2.8}$~cm$^{-3}$, and the vertical dotted line 
the critical density of the ionic specie, if within the range displayed. 
The best fit inner radius is r$_{\rm in}=$10$^{21.4}$~cm (green symbols).
}
\label{fig:hiidiag}
\end{figure}

The results of the computed grid are displayed in Figure~\ref{fig:hiidiag}. 
It shows the intensity predictions of the compact model labeled $\mathscr{C}$ 
for each ionic line as a function of density, n$\rm{_{H}}$, and inner radius, r$\rm{_{in}}$, 
for a starburst age of 3.7~Myr (age of the best fit model, see Sect.~\ref{sect:age}). 
In this section, we discuss the influence of each grid parameter 
on the model predicions and possible solution values. 
These parameters are linked via the quantity $U$, which is the ionisation parameter, 
defined as the dimensionless ratio of the incident ionising 
photon surface density to the hydrogen density n$\rm{_{H}}$:
$U = \frac{Q(H)}{4\pi r_{\rm in}^{2}n_{H}c}$,
where $Q(H)$ is the rate of hydrogen-ionising photons
striking the illuminated face and $c$ is the velocity of light. 
$U$ describes how many ionising photons arrive per atom.

\subsubsection{The density of the gas}
\label{sect:hden}
In general terms 
the predicted line intensities are dependent on the gas density through 
both the critical density of each atom, and the ionisation parameter, $U$. 
Line intensities increase with increasing gas density until the critical 
density of a fine-structure line is reached. 
Above such critical density, the de-excitation process is 
dominated by collisions rather than radiative decay.
Therefore the maximum gas density is well determined 
by the critical density of the brightest line tracers. 
The gas density also plays an important role in determining the 
population of the different ionisation stages. 
The ionisation parameter, $U$, is inversely proportional to the density 
(bottom right panel of Fig.~\ref{fig:hiidiag}). 

We find that the minimum $\chi^2$ occurs for a solution with a density 
around 10$^{2.8}$~cm$^{-3}$ (dashed vertical line in Fig.~\ref{fig:hiidiag}). 
This derived gas density compares well with density diagnostics proposed 
in the literature, for example, based on the ratio of the [S~{\sc iii}] lines at 
18.7 and 33.5~$\mu$m \citep{houck-1984}. This ratio is sensitive to the 
density since these lines originate from the same ion but with significantly 
different critical densities (Table~\ref{table:lines}).
From \cite{abel-2005} and requiring 
$\mathrm{\frac{[S~III]~18~\mu\,m}{[S~III]~33~\mu\,m}\sim1.0}$, 
we obtain a density of the ionised gas of $\rm{n_{H} = 10^{2.7}~cm^{-3}}$.
The value of the electron density predicted by the diagnostic plot 
in \cite{houck-1984} is also around $\rm{10^{2.8}~cm^{-3}}$.
Following the method from \citet*{kingsburgh-1992,kingsburgh-1994}
who use the programs $EQUIB$ and $RATIO$ (developed at UCL by I. D. Howarth 
and modified by S. Adams), we also find $\rm{n_{H}\sim10^{2.8}~cm^{-3}}$.

\subsubsection{The age of the burst}
\label{sect:age}
\begin{figure}[!htp]
\centering
\includegraphics[width=8.8cm]{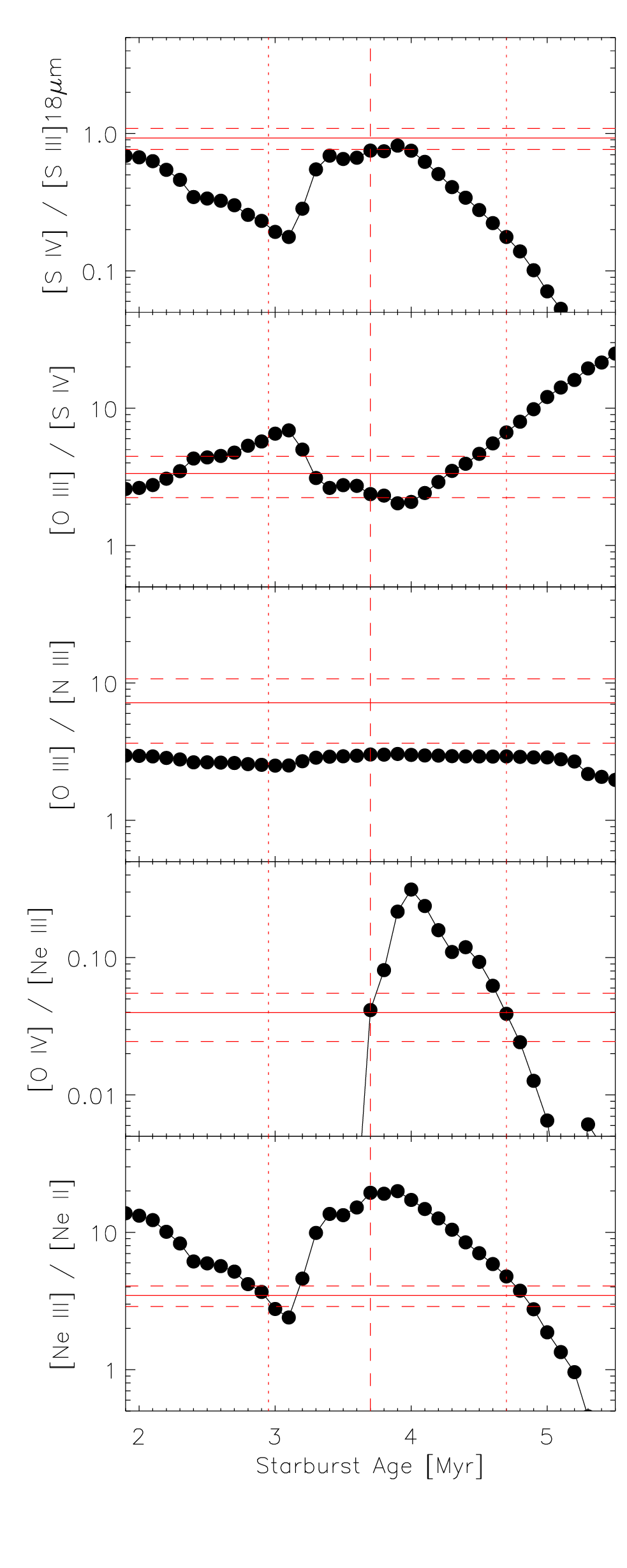}
\caption{Model predicted intensity ratios as a function of burst age A$\rm{_{burst}}$, 
for a hydrogen density of $\rm{10^{2.8}~cm^{-3}}$, and inner radius of 10$^{21.4}$~cm. 
The true burst age of these models is an average of A$\rm{_{burst}}$ and the older 
stellar population of 9.5~Myr, although the latter has a negligible influence 
on the high-ionisation lines considered here. 
The horizontal solid lines represent the observed ratios and the dashed lines 
the uncertainties on the ratios. 
The vertical dashed line shows the solution burst age of 3.7~Myr, 
and the vertical dotted lines the age of 3~Myr and 4.7~Myr that would 
better match the [Ne~{\sc iii}]/[Ne~{\sc ii}] ratio.
}
\label{fig:hiiratios}
\end{figure}

The spectrum of the starburst (intensity and shape) changes drastically 
with time in the age range considered here. 
The age of the burst sets the hardness of the incident radiation field.
Figure~\ref{fig:hiiratios} shows line ratios of [Ne~{\sc iii}]/[Ne~{\sc ii}], 
[O~{\sc iv}]/[Ne~{\sc iii}], [O~{\sc iii}]/[N~{\sc iii}], [O~{\sc iii}]/[S~{\sc iv}], 
and [S~{\sc iv}]/[S~{\sc iii}]~18.7~$\mu$m as a function 
of burst age. 
The observed ratios agree with an age of burst of 3.5-5~Myr.
These values are also in agreement with the age of 3.5~Myr for the 
brightest knot B and 9~Myr for knot C, and a mixture of very young 
and older bursts, found in \cite{adamo-2010}. 
After 5~Myr, the ionising part of the burst spectrum decreases, 
the high ionisation lines start to disappear, and in particular 
the [Ne~{\sc iii}]/[Ne~{\sc ii}] ratio drops off (bottom panel of Fig.~\ref{fig:hiiratios}). 
This is why we do not compute models with A$\rm{_{burst} > 5~Myr}$ in our grid. 
The best fit on the neon lines favours a relatively young burst, of about 3~Myr. 
However, the emission of the highest excitation potential [O~{\sc iv}] line (54.9~eV) 
is rather favoured by models with ages between 3.7 and 4.7~Myr.
Indeed this time frame corresponds to the time needed for the onset of Wolf-Rayet stars. 
A burst age of 4.7~Myr is also a good match to the [Ne~{\sc iii}]/[Ne~{\sc ii}] ratio, 
although it does not agree with the intensity of the [S~{\sc iv}] line, which 
is one of the lines that has the highest weight in the $\chi^2$. 
The best single solution model in that case has an age of burst of 3.7~Myr.

\subsubsection{The inner radius}
\label{sect:rin}
All lines are sensitive to the inner radius $\rm{r_{in}}$. 
Its choice has no direct influence on the hardness of the radiation field, 
but rather on its intensity. With increasing radius, the total input luminosity 
is distributed over a larger surface, decreasing the number of ionising 
photons per surface area, and therefore the ionisation rate.
$U$ decreases as the inverse square of the radius. 
Intensities of species with a high ionisation potential, such as [O~{\sc iv}], 
decrease with increasing $\rm{r_{in}}$, and the inverse is observed 
for low ionisation species such as [N~{\sc ii}].
For r$\rm{_{in} < 10^{21.0}~cm}$, atoms are preferably present in high ionisation states.
Models with such small r$\rm{_{in}}$ often predict too much [O~{\sc iv}] and [S~{\sc iv}], 
and not enough [Ne~{\sc ii}] and [S~{\sc iii}]. 
[Ne~{\sc iii}] and [S~{\sc iv}] intensities are constant at low r$\rm{_{in}}$ 
and then drop with increasing r$\rm{_{in}}$.
[S~{\sc iii}], [N~{\sc iii}] and [O~{\sc iii}] intensities increase with low 
r$\rm{_{in}}$ ($<$~10$^{21.5}$ cm) because of an ionisation balance 
between S$^{2+}$, N$^{2+}$, O$^{2+}$ and S$^{3+}$, N$^{3+}$, O$^{3+}$.  
Then their intensities drop with increasing r$\rm{_{in}}$.
The opposite effect is observed for the lower ionisation state species 
such as [Ne~{\sc ii}] and [N~{\sc ii}] which increase with r$\rm{_{in}}$.

The observations are best matched for an inner radius of 10$^{21.4}$~cm (800~pc).
At that distance, the model cloud is a very thin shell of large radius, 
at the center of which resides the central starburst.
The resulting geometry is plane-parallel.
This is not a realistic geometry, as discussed in Sect.~\ref{sect:limits}, 
but we do not aim to model the spatial structures in Haro\,11.

\subsubsection{Parameters of the best fit model $\mathscr{C}_{HII}$}
\begin{figure}[!htp]
\centering
\includegraphics[width=8.8cm]{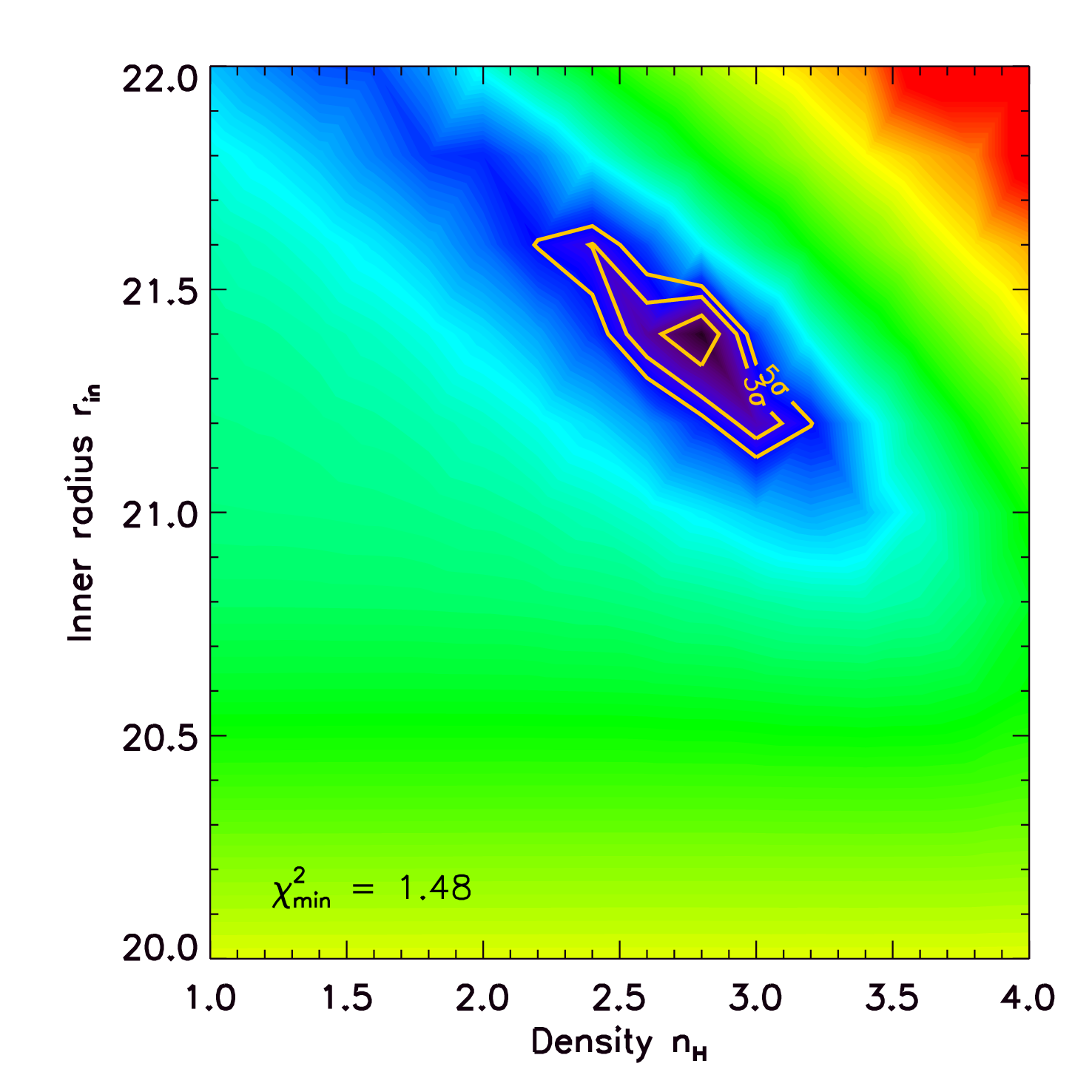}
\includegraphics[width=8.8cm]{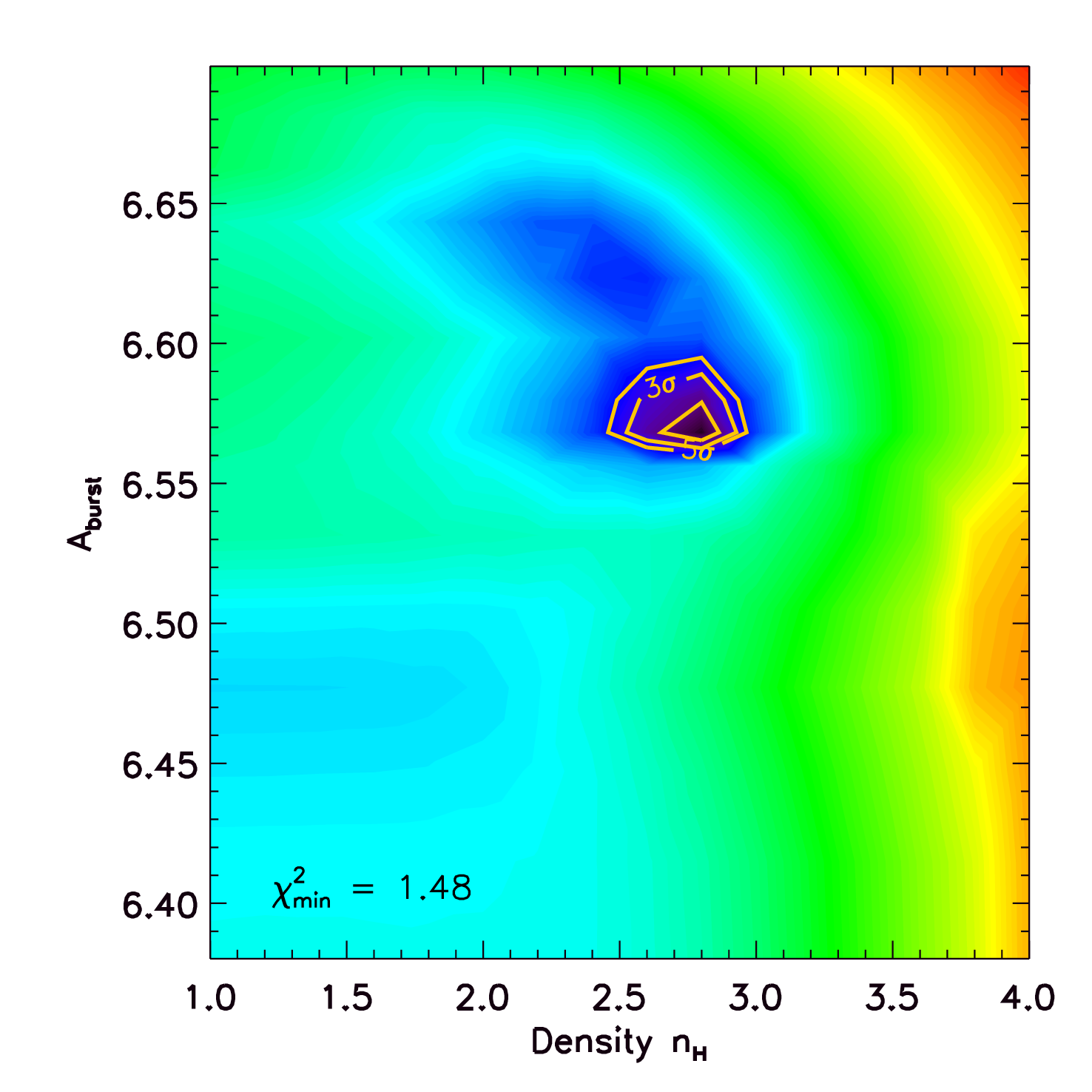}
\caption{
Contour plots of the $\chi^2$ values of the 3D grid (log scale). 
The parameters are given in logarithmic values. 
Units are cm$^{-3}$ for the density, yr for the burst age, and cm for the inner radius.
The minimum reduced $\chi^2$ found for $\mathscr{C}_{HII}$ is $\chi^2_{min}=1.48$. 
Contour levels at 1-$\sigma$, 3-$\sigma$, and 5-$\sigma$ 
from the minimum $\chi^2$ are displayed in orange. 
There are 5 other models within 3-$\sigma$ of $\chi^2_{min}$. 
The red color on the plots correspond to $>$1000-$\sigma$ level. 
{\it Top:} $\chi^2$ contours in the $\rm{n_{H}}$-$\rm{r_{in}}$ plane, 
for $\rm{A_{burst} = 3.7~Myr}$ (or 6.57 in log). 
{\it Bottom:} $\chi^2$ contours in the $\rm{n_{H}}$-$\rm{A_{burst}}$ plane, 
for $\rm{r_{in} = 10^{21.4}~cm}$. 
}
\label{fig:chi2}
\end{figure}

We have minimised the $\chi^2$ to find the parameters of the 
best fit model for the compact H~{\sc ii} region described above. 
Figure~\ref{fig:chi2} shows contour plots of the reduced $\chi^2$ 
computed for our grid of models. 
Contour levels at 1-$\sigma$, 3-$\sigma$, and 5-$\sigma$ 
from the minimum $\chi^2$ are displayed in orange. 
The intensities of the spectral lines included in the fit are best reproduced 
with the parameters: $\rm{n_{H} = 10^{2.8}~cm^{-3}}$, 
$\rm{A_{burst} = 3.7~Myr}$, and $\rm{r_{in} = 10^{21.4}}$~cm. 
The ionisation parameter is $\rm{U=10^{-2.5}}$. 
The reduced $\chi^2$ of this model solution, which has the minimum $\chi^2$, is 1.48. 
This value is close to 1, indicating that this model describes the data well, 
with a probability of $\sim$70\%. 
For clarity throughout the following Sections, we will refer to this model 
solution of the compact H~{\sc ii} region as $\mathscr{C}_{HII}$.

With $\mathscr{C}_{HII}$, the strongest MIR line, [Ne~{\sc iii}], is reproduced 
within 15\%, as well as the [O~{\sc iv}] line.
The two [S~{\sc iii}] lines are achieved within 5\% and their ratio within 1\%.
The modeled [S~{\sc iv}] and [Ar~{\sc iii}] lines deviate by 20\%.
The [N~{\sc iii}] line is reproduced within 40\%. 
The [Ne~{\sc ii}] line, however, is under-predicted by a factor of 5 
by $\mathscr{C}_{HII}$.

We then compare the line intensities predicted by $\mathscr{C}_{HII}$ 
to the PACS lines, [O~{\sc iii}] and [N~{\sc ii}]~122~$\mu$m, 
which were not used to constrain the model fit since they can originate from 
less dense ionised gas.
We find that [O~{\sc iii}] is under-predicted by a factor of 2, which is within 
our  expectations, while the [N~{\sc ii}]~122~$\mu$m is under-predicted by a factor of 4. 
The dense H~{\sc ii} region is not the primary source of [N~{\sc ii}]. 
The prediction for the [N~{\sc ii}]~205~$\mu$m line falls below its derived upper limit.

We discuss the origin of the [Ne~{\sc ii}] and [N~{\sc ii}] lines from diffuse 
low-ionisation gas in Sect.~\ref{sect:diffus} and the possible origin of the 
[O~{\sc iii}] emission from diffuse high-ionisation gas in Sect.~\ref{sect:diffus}.

We also find that $\mathscr{C}_{HII}$ contributes very little 
to the [O~{\sc i}] emission (3\%), as expected since the atomic oxygen 
is a tracer of the neutral gas. $\mathscr{C}_{HII}$ shows very little contribution 
to the [C~{\sc ii}] ($<$1\%) and [Si~{\sc ii}] (11\%) lines as well.
The energy required to create C$^{++}$ ions is 24.4~eV, close to that 
of S$^{++}$, therefore when the radiation field is sufficiently hard, as is 
the case inside the H~{\sc ii} region, C is doubly rather than singly ionised.

\subsection{The dense photodissociation regions: $\mathscr{C}_{PDR}$}
\label{sect:pdr}
We subsequently investigate the properties of the neutral gas and compare model 
predictions to the [C~{\sc ii}] and [O~{\sc i}]~63 and 145~$\mu$m lines. 
To this end, we extend $\mathscr{C}_{HII}$ into the atomic 
phase and the outer envelope of the molecular phase, until the 
molecular fraction H$\rm{_{2}}$/H reaches 99\%. 
The modeled shell is fully covered by the gas. 
The ionised cloud, with average density of 10$^{2.8}$ cm$\rm{^{-3}}$, 
transitions from the H~{\sc ii} region into a dense PDR, 
which is where the [O~{\sc i}] originates.
This transition between these two phases in a constant pressure model 
translates into a jump of the gas density at the ionisation front.
With this model, the average density of the PDR is 10$^{5.1}$ cm$\rm{^{-3}}$. 
The radiation field striking the PDR is: $\rm{G_{0} = 3.5 \times 10^{3}}$ 
Habing unit ($\rm{1.6 \times 10^{-3} ergs~cm^{-2}~s^{-1}}$). 
The heating of the gas in the PDR is dominated by grain photoelectric heating (85\%), 
and there is a second order contribution from collisions with H$\rm{^0}$, He, and H$\rm{_2}$.
The cooling is done by radiative de-excitation mainly of the [O~{\sc i}]~63~$\mu$m (65\%), 
[Si~{\sc ii}] (20\%), [C~{\sc ii}]~157~$\mu$m and [O~{\sc i}]~145~$\mu$m ($<$7\% each) lines. 
At such low A$_v <$1, the [O~{\sc i}] lines are optically thin.

The [O~{\sc i}] 63 and 145~$\mu$m absolute intensities are over-estimated 
by a factor of 10, while their ratio is reproduced within 15\% by this dense PDR component.
Along with the [O~{\sc i}]~63~$\mu$m line, [C~{\sc ii}]~157~$\mu$m and 
[Si~{\sc ii}]~35~$\mu$m are the brightest PDR lines. 
While the [Si~{\sc ii}] line intensity is over-predicted by a factor of 2 
by the dense model, the [C~{\sc ii}] line agrees with the observation within 10\%. 

The consistent over-prediction of the two [O~{\sc i}] lines may be explained by 
the fact that, in reality, the dense PDR does not cover the full dense H~{\sc ii} region. 
Assuming that all of the [O~{\sc i}] arises from a single dense phase 
(and not the diffuse gas, see Sect.~\ref{sect:ciidiff}), we reduce the covering factor 
of the dense PDR to only 10\%, which is a proxy for dense clumps 
\citep{hunter-2001,groves-2008}. 
The PDR model with a covering factor of 1/10 matches the observed [O~{\sc i}] emission, 
but under-estimates the [C~{\sc ii}] emission by an order of magnitude. 
The main conclusion from this is that the dense PDR component cannot account 
for the bulk of the [C~{\sc ii}] emission observed in Haro\,11, but only for 10\%. 
This mismatch for the [C~{\sc ii}] is due to the relatively low critical density of [C~{\sc ii}] 
($\rm{2 \times 10^{3}~cm^{-3}}$) compared to the high density reached in the PDR 
($\rm{n_{PDR} = 10^{5.1}~cm^{-3}}$), generated by the pressure equilibrium 
and the drop of gas temperature between the ionised and neutral phases. 
Indeed, Cloudy PDR models with constant densities of $\sim$10$^{3.5}$~cm$^{-3}$ 
and a covering factor of unity would reproduce the absolute intensity 
of the two [O~{\sc i}] lines simultaneously, and yet under-predict the 
[C~{\sc ii}] by a factor of 2. Moreover, the H~{\sc i} mass of such models would 
exceed the upper limit of $\rm{10^{8}~M_{\odot}}$ found by \cite{bergvall-2000}.

The small fraction of [C~{\sc ii}] that originates from the dense PDR is at odds with the 
finding of \citet{bergvall-2000}, who estimate that more than 80\% of the observed 
[C~{\sc ii}] in Haro\,11 comes from the neutral gas. They use stellar clusters of 
T$\rm{_{eff}} \sim$~35~000-40~000~K to account for the [O~{\sc iii}]~88~$\mu$m 
emission, which contributes less than 20\% to the [C~{\sc ii}] emission, and 
they assume that the rest of the [C~{\sc ii}] ($>$80\%) is associated with PDRs. 
Then, they find $\rm{n_{PDR} = 2 \times 10^{3}~cm^{-3}}$ using the \cite{kaufman-1999} 
model, which assumes a constant density. 
The much higher density that we find is compatible with the findings of \cite{vasta-2010} 
who analysed ISO observations with PDR models for 
several dwarf galaxies, including Haro\,11, and find PDR densities 
between $\rm{10^{4}}$ and $\rm{9 \times 10^{4}~cm^{-3}}$, and G$_0$ between 60 and 
$\rm{8 \times 10^{2}~Habing}$. 
They also estimate that 10 to 60\% of [C~{\sc ii}] comes from the ionised gas, 
which is in agreement with the diffuse ionised model contribution to the 
[C~{\sc ii}] line that we discuss in Sect.~\ref{sect:ciidiff}.

We have also explored the influence of cloud geometry on the line predictions 
by using the PDR model Kosma-$\tau$ \citep{roellig-2006}, 
in which the illumination of the cloud is external, simulating a spherical cloud structure. 
At corresponding values of $\rm{n_{H}}$ (10$^{5.1}$~cm$\rm{^{-3}}$) and 
G$_0$ ($\rm{3.5 \times 10^{3}}$~Habing) found with Cloudy, 
Kosma-$\tau$ predicts ratios of [O~{\sc i}]~145~$\mu$m/[C~{\sc ii}]$\sim$1 
and [O~{\sc i}]~145/63~$\mu$m$\sim$0.05. 
The [O~{\sc i}]~145~$\mu$m/[C~{\sc ii}] model ratios are over-predicted, 
as found with Cloudy, compared to our observed value of 0.08.  
However, if we attribute only 10\% of the [C~{\sc ii}] emission to the PDR, then 
[O~{\sc i}]~145~$\mu$m/[C~{\sc ii}]$\rm{_{PDR}}\sim$~0.1 from Kosma-$\tau$, 
which is in line with our observations.
The spherical geometry affects both [O~{\sc i}] and [C~{\sc ii}] lines similarly to first order, 
and is not the primary reason for the discrepancy between observed and predicted 
[C~{\sc ii}] intensities.

We prefer to impose the pressure equilibrium between the compact 
H~{\sc ii} region and the dense PDR, because this gives physical insight 
into the presence of dense but fragmented PDRs \citep[e.g.][]{hunter-2001}. 
Although the pressure equilibrium assumption is not valid 
on large scales (nor is the assumption of a constant density) and, 
in particular, is expected to break in molecular clouds, 
here we do not model the molecular phase. 
We are interested in the interface between the dense H~{\sc ii} region 
and the PDR, the presence for which there is observational evidence. 
This model solution of the dense PDR with covering factor of 10\% 
is noted $\mathscr{C}_{PDR}$.
Next, we will discuss possible ways to lower the density in the PDR 
to account for the [C~{\sc ii}] emission, by either combining a high density with 
a low density model (Sect.~\ref{sect:ciidiff}), representative of the multi-phase ISM 
of galaxies, or by including a magnetic field (Sect.~\ref{sect:bmag}).

\section{A model for the diffuse medium: $\mathscr{D}$}
\label{sect:diffus}

\subsection{Need for a softer radiation field}
The major discrepancies between observations and model predictions 
in the ionised gas are seen for the [Ne~{\sc ii}]~12.6~$\mu$m and 
[N~{\sc ii}]~122~$\mu$m lines, which are both under-predicted by a 
factor of $\sim$5 in the dense H~{\sc ii} region model $\mathscr{C}_{HII}$, 
which is not unexpected. 
With excitation potentials of 14.5 and 21.6~eV respectively, emission 
of N$^+$ and Ne$^+$ originates only from the ionised gas, from both 
the dense H~{\sc ii} region and the diffuse medium. 
For [N~{\sc ii}]~122~$\mu$m, $\mathscr{C}_{HII}$ does not reproduce 
the observed value even when varying the 3 free parameters. Only 
$\rm{r_{in} > 10^{21.8}}$~cm may agree (bottom left panel of Fig.~\ref{fig:hiidiag}). 
The contribution of the compact H~{\sc ii} region to this line is thus marginal; 
mostly because the radiation field produced by the young starburst is too hard 
and favors the ionisation of N$^{++}$ rather than N$^+$. 
For the [Ne~{\sc ii}] line, a small change in the age of the burst would 
impact on its prediction, as discussed in Sect.~\ref{sect:age}.
For the same values of density and inner radius, a model of age 3~Myr 
(instead of 3.7~Myr) would agree better with the observed intensity of [Ne~{\sc ii}]. 
For a burst age of 3.7~Myr, higher densities ($\rm{n_{H} > 10^{3}~cm^{-3}}$) 
and inner radii ($\rm{r_{in} > 10^{21.6}}$~cm) are required to reproduce 
the observed [Ne~{\sc ii}] (top right panel of Fig.~\ref{fig:hiidiag}). 
However, such values of $\rm{n_{H}}$ and $\rm{r_{in}}$ would no longer 
agree with the other ionic lines. 
In order to reconcile the modeled [Ne~{\sc ii}] and [N~{\sc ii}] with 
their observations, a component with softer radiation field is required. 
This highlights the fact that one component is not enough the account 
for the emission of all the observed lines due to the presence 
of several phases in the ISM of galaxies with different properties. 
\revised{
Our physical picture of the ISM of dwarf galaxies is of a highly permeable 
medium, where some of the ionising photons from the central starburst 
escape the H~{\sc ii} region and travel large distances. 
This means that in dwarf galaxies, larger effective volumes can be maintained 
in an ionised state, compared to those of dusty starbursts. 
}

To model this additional ionised component we have tried three different methods: 
(1)~set the cloud further away (higher r$\rm{_{in}}$) under the same starburst conditions, 
and allowing the density to be a free parameter, 
(2)~stop $\mathscr{C}_{HII}$ before it reaches the ionisation front 
and take the output spectrum as the input spectrum of a lower density medium, 
(3)~opt for a softer radiation field by including a scaled local interstellar radiation field. 
However, all of these methods fail to reproduce the observed line intensities. 
Either the radiation is still too hard (case (1) and (2) when stopping too early), 
or too soft (case (3) and (2) when stopping too close to the ionisation front).

The radiation field that works best to explain the observed [Ne~{\sc ii}] 
and [N~{\sc ii}] lines is a representative stellar SED from the Kurucz library 
of relatively low temperature (see next subsection). 
This is supported by the fact that more evolved stars producing low energy 
photons are present in Haro\,11 \citep{micheva-2010} but not accounted for 
in the compact model $\mathscr{C}$ ($\mathscr{C}_{HII}$+$\mathscr{C}_{PDR}$). 
This older population affects mainly the emission of [N~{\sc ii}], [Ne~{\sc ii}], 
and to a lesser extent [S~{\sc iii}].

\subsection{Properties of the model $\mathscr{D}$}
\begin{figure}[!htp]
\centering
\includegraphics[width=8.8cm]{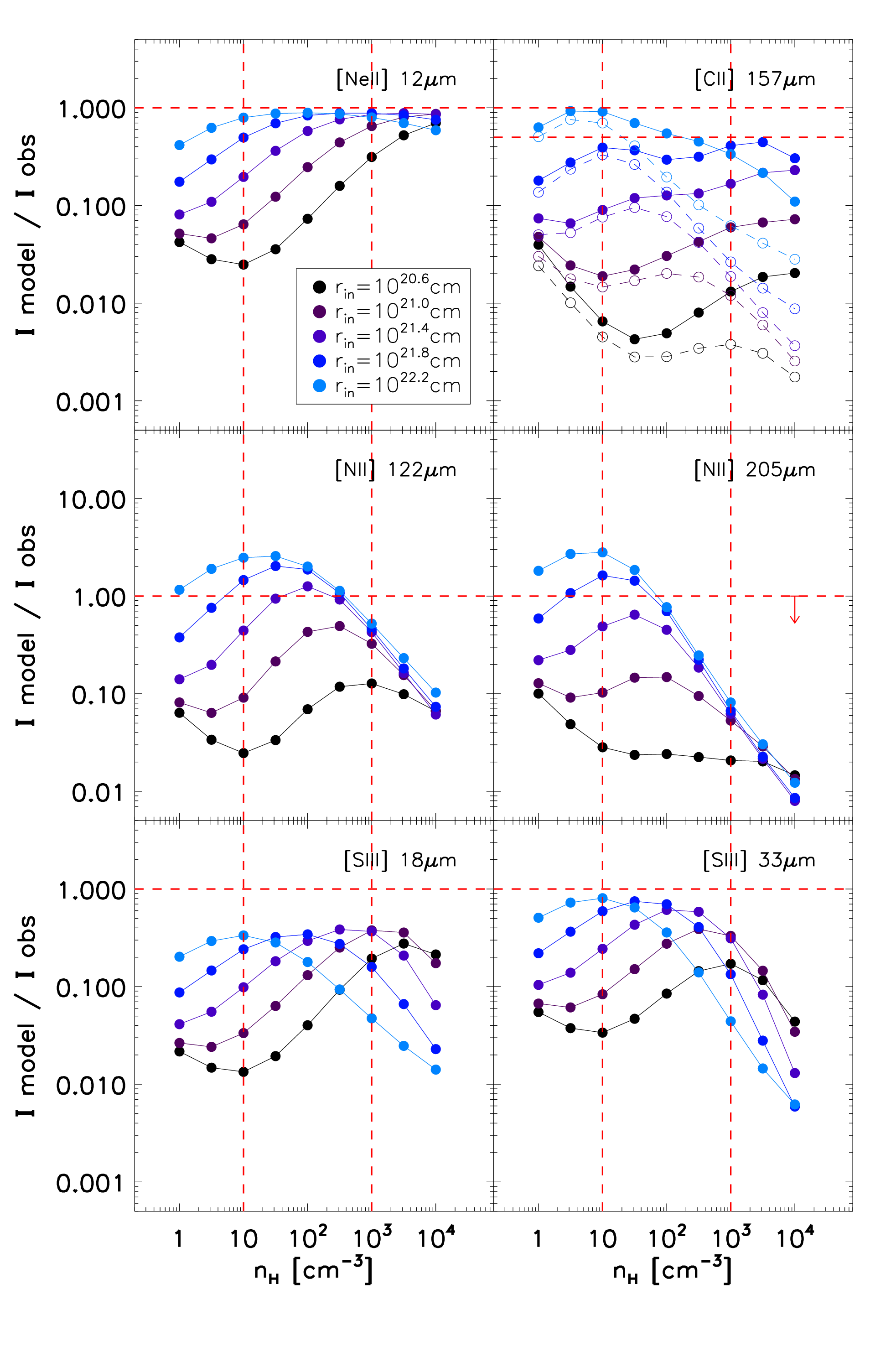}
\caption{
Predictions for the model $\mathscr{D}$ of the [Ne~{\sc ii}], [C~{\sc ii}], 
[N~{\sc ii}], and [S~{\sc iii}] line intensities (I$_{model}$), normalised by 
their observed value (I$_{obs}$) as a function of density $\rm{n_{H}}$.
Models are for a diffuse ionised/neutral gas, with constant density, stellar 
temperature of 35~000~K, and are stopped when molecules start to form. 
The [S~{\sc iii}], [Ne~{\sc ii}] and [N~{\sc ii}] lines emit in the ionised phase. 
The contribution of the ionised phase to the [C~{\sc ii}] line is indicated 
with open circles. 
The vertical dashed lines indicate density values of the diffuse models 
$\mathscr{D}_{l}$ ($\rm{n_{H} = 10~cm^{-3}}$) and $\mathscr{D}_{h}$ 
($\rm{n_{H} = 10^{3}~cm^{-3}}$) defined in Sect.~\ref{sect:ciidiff}.
}
\label{fig:diffus}
\end{figure}

We ran constant density models of diffuse ionised gas, stopped at the ionisation 
front, with stellar temperature between 20~000 and 40~000~K. 
The best fit to the [N~{\sc ii}]~122~$\mu$m and [Ne~{\sc ii}] emission is obtained 
for a stellar temperature of 35~000~K. Figure~\ref{fig:diffus} shows how the 
[N~{\sc ii}], [Ne~{\sc ii}], [S~{\sc iii}], and [C~{\sc ii}] lines vary with density for 
several values of $\rm{r_{in}}$. We refer the reader to the Sections \ref{sect:hden} 
and \ref{sect:rin} for an interpretation of the behavior of the lines. 
The range of possible solutions for the [N~{\sc ii}]~122~$\mu$m emission are densities 
below $\rm{10^{3}~cm^{-3}}$ and $\rm{r_{in}}$ higher than $\rm{10^{21.0}~cm}$. 
[Ne~{\sc ii}] also agrees well with these conditions.
At T$\rm{_{eff}}$ of 35~000~K, [S~{\sc iii}] also emits, but remains below its observed value. 
Very little [N~{\sc iii}], [Ne~{\sc iii}], [O~{\sc iii}], and [S~{\sc iv}] are produced 
in this diffuse component, and no [O~{\sc iv}].
Although [N~{\sc ii}]~122~$\mu$m has a relatively low critical density, 
the density of the N$^+$ emitting medium is usually well constrained by the 
[N~{\sc ii}]~205~$\mu$m line. Unfortunately, the [N~{\sc ii}]~205~$\mu$m line 
is difficult to observe with PACS because of spectral leakage (see Sect.~\ref{sect:pacs}). 
The upper limit on the [N~{\sc ii}]~205~$\mu$m line sets an upper limit on the ratio 
[N~{\sc ii}]~205/122~$\mu$m of 0.75 which corresponds to a lower limit on the density 
of the ``diffuse'' gas component of 10~cm$^{-3}$ \citep[e.g.][]{rubin-1994,oberst-2006}. 
The [N~{\sc ii}]~122~$\mu$m line critical density sets an upper limit of $\rm{10^{3}~cm^{-3}}$.
Models with densities between 10 and $\rm{10^{3}~cm^{-3}}$ work equally 
well for the ``diffuse'' medium in acounting for the [N~{\sc ii}]~122~$\mu$m 
and [Ne~{\sc ii}] emission.

\subsection{[C~{\sc ii}] contribution from the model $\mathscr{D}$}
\label{sect:ciidiff}
Since $\mathscr{C}_{PDR}$ predicts very little [C~{\sc ii}]~157~$\mu$m, 
we expect a significant contribution of the diffuse medium to the C$^{+}$ emission. 
C$^{+}$ is usually found in the surface layers of far-UV 
illuminated PDRs \citep{stacey-1991,negishi-2001}, 
but it can also come from the ionised gas, with a contribution up to 50\%
\citep{madden-1993,carral-1994,heiles-1994,abel-2005}.
With ionisation potentials lower than that of hydrogen, both 
C$^+$ and Si$^+$ can be produced by low energy photons 
or excited by collisions with electrons or hydrogen atoms, depending 
on the degree of ionisation from the emitting medium. 

Figure~\ref{fig:diffus} shows the predicted [C~{\sc ii}] line intensities from the diffuse 
medium with a stellar temperature of 35~000~K, as a function of gas density. 
The contribution of the diffuse phase (ionised+neutral) is in filled circles, and 
the contribution of the ionised phase only shown by open circles. 
Calculations are stopped when molecules start to form and the fraction 
of $\rm{H_{2}}$/H reaches 10\%.
At low $\rm{n_{H}}$, the empty and filled circles are very close: a large fraction of the 
[C~{\sc ii}] can arise from the ionised phase, until [C~{\sc ii}] is collisionally de-excited 
by e$^-$ (n$_{crit}$~=~50~cm${\rm ^{-3}}$), 
and emits mostly in the neutral phase, its intensity then increasing with density. 
At $\rm{r_{in} = 10^{22.2} ~cm}$, the [C~{\sc ii}] predictions decrease for 
$\rm{n_{H} > 10^{2}~cm^{-3}}$, because the atomic phase is very thin as 
the material is cold and enters quickly into the molecular phase where the 
models stop. We disregard these models which do not predict enough [C~{\sc ii}]. 
The range of possible solutions for the [C~{\sc ii}]~157~$\mu$m emission are 
$\rm{r_{in}}$ higher than $\rm{10^{21.4} ~cm}$, and are degenerate in $\rm{n_{H}}$. 
We can set a low-density case where $\rm{n_{H} = 10~cm^{-3}}$ -- for example. 
Then $\rm{r_{in} = 10^{21.8} ~cm}$ is the best fit, and [C~{\sc ii}] comes mostly from 
the ionised phase, accounting for $\sim$45\% of the observed value. 
We also consider a high-density case where $\rm{n_{H} = 10^{3}~cm^{-3}}$, 
and $\rm{r_{in} = 10^{21.8} ~cm}$. In this case, [C~{\sc ii}] is mostly emitted in the 
neutral phase, accounting for $\sim$40\% of the observed value. 
Both models fill a larger volume than the compact model $\mathscr{C}$.

\subsubsection{Constraint from the H~{\sc i} 21-cm line}
The H~{\sc i} 21-cm mass upper limit of $\rm{10^{8}~M_{\odot}}$
from \cite{bergvall-2000} is an important gas diagnostic of the diffuse 
atomic gas and brings a strong constraint that we can use to 
differentiate between the low-density and high-density case models. 
The H~{\sc i} mass from $\mathscr{C}_{PDR}$ is $\rm{10^{7}~M_{\odot}}$. 
While the models from Fig.~\ref{fig:diffus} were stopped at a fraction 
of $\rm{H_{2}}$/H reaching 10\% to compute entirely the ionised 
and atomic phases, most of those models exceed the H~{\sc i} mass 
upper limit.
By stopping those models when the upper limit on the H~{\sc i} mass 
is reached, we find that the low-density case model contribution to the 
[C~{\sc ii}] line is unchanged since it comes from the ionised phase, 
while the high-density case model contributes to only $\sim$20\% of 
the [C~{\sc ii}] line intensity, since the atomic phase is stopped at lower 
A$_V$ than previously computed.

For the rest of this study, we refer to the low-density case model as 
$\mathscr{D}_{l}$, and to the high-density case model as $\mathscr{D}_{h}$. 
We cannot exclude the presence of a diffuse neutral gas, but in all cases, 
there is a prominent contribution to the [C~{\sc ii}] from the ionised gas. 
We estimate that $\sim$40\% originates in $\mathscr{D}_{l}$; 
the rest coming from $\mathscr{C}_{PDR}$ and little from $\mathscr{D}_{h}$.
Moreover, the diffuse models do not contribute more than 15\% to the [O~{\sc i}] lines, 
which are reliable tracers of the PDR.

\section{Influence from other possible components}
\label{sect:disc}

\subsection{Magnetic fields}
\label{sect:bmag}
Magnetic fields impact the model solution, dominating the pressure 
deep into the cloud, and they are expected to be important in star-forming 
regions \citep{shaw-2009}. 
\cite{robishaw-2008} found magnetic fields on the order of a few mG in ULIRGs, 
from OH Zeeman splitting.
In local dwarf irregular galaxies, magnetic fields are found to be weak, of a few 
$\mu$G to 50~$\mu$G in more extreme cases 
\citep[e.g. NGC~1569, NGC~4214;][]{chyzy-2011,kepley-2010,kepley-2011}.
No study on magnetic fields has been conducted for Haro\,11, 
so we did not include them in our grid calculation. 
Nevertheless, in view of the fact that we have found that the density 
of the dense PDR plays a determining role in the predicted line fluxes, 
we investigate the influence of adding a magnetic field 
$B_{mag}$ of strength 1~$\mu$G to 3~mG to the solution model 
$\mathscr{C}$ described in Sect~\ref{sect:results}.

\begin{figure}[!htp]
\centering
\includegraphics[height=10cm,width=8.8cm]{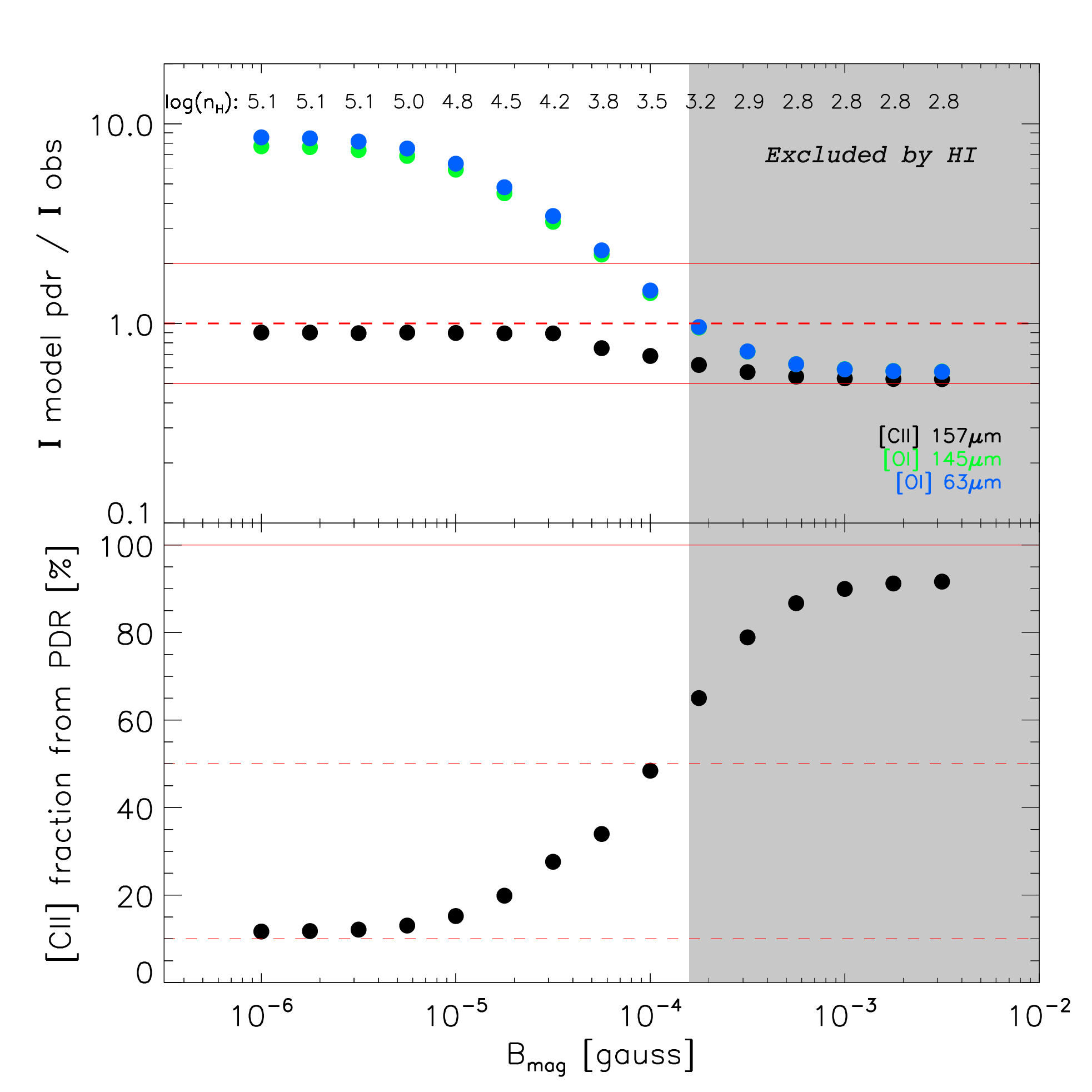}
\caption{
Influence of the magnetic field on the PDR lines. 
{\it Top}: Model predictions of the [C~{\sc ii}] and [O~{\sc i}] intensities 
(I$_{model}$) normalised by their observed value (I$_{obs}$) 
as a function of magnetic strength $B_{mag}$.
The effective density in the PDR is indicated for each $B_{mag}$ value. 
{\it Bottom}: Fraction of [C~{\sc ii}] coming from the PDR, scaled to 
the [O~{\sc i}] 145~$\mu$m prediction, assuming that all of it originates in the PDR. 
Models in the shaded area are not valid with respect to the 
upper limit on the H~{\sc i} column density.
}
\label{fig:bmag}
\end{figure}

\revised{
The magnetic field ($B_{mag}$) is expressed as a pressure term ($P_{mag}$) 
through the cloud by the following equation: 
\begin{equation}
P_{mag} = \frac{B^{2}_{mag}}{8\pi} ~\mathrm{and}~ 
 B_{mag}(r) = B_{mag}(0) \left( \frac{n_{H}(r)}{n_{H}(0)} \right) ^{\kappa} 
 ~\mathrm{, where}~ \kappa = 2/3
\end{equation}
where $r$ is the depth of the cloud, $B_{mag}(0)$ and $n_{H}(0)$ are 
the initial values of the magnetic field and hydrogen density. 
This density--magnetic field relationship from \cite{crutcher-1999} corresponds to 
the case where gravity dominates over magnetic support in the cloud collapse, and 
the magnetic flux is conserved. 
}
Figure~\ref{fig:bmag} shows the predicted [C~{\sc ii}] and [O~{\sc i}] intensities 
with increasing $B_{mag}$. For low values of $B_{mag}$, thermal pressure, 
expressed as $P_{gas}~\propto~n_{H}~T_{gas}$, dominates 
and the densities in the PDR are high, with strong emission of [O~{\sc i}], as seen previously. 
With increasing $B_{mag}$, magnetic pressure starts to dominate over gas pressure 
and the transition into the PDR is smooth, resulting in lower densities in the PDR, and 
less emission of [O~{\sc i}]. The [C~{\sc ii}]/[O~{\sc i}] ratio increases 
with increasing $B_{mag}$. 
The [C~{\sc ii}] prediction falls between a factor of 0.5 and 1 of its observed value 
for all models. 
Values of $B_{mag}$ up to $\rm{10^{-3.75}~G}$ predict enough 
[O~{\sc i}] emission.
For $B_{mag}$ $\rm{> 10^{-3.75}~G}$, the models under-predict 
the intensity of the PDR lines, which would require extra input power, 
although we are still within the errors (factor of 2). 

However, when lowering the density, the size of the PDR layer increases, 
and the H~{\sc i} mass as well. 
With no magnetic field, we find M(H~{\sc i})$\rm{= 10^{7}~M_{\odot}}$ with 
a covering factor of $\sim$10\%. For $B_{mag} > 10^{-4}$~G, we find 
M(H~{\sc i})$\rm{> 2 \times 10^{8}~M_{\odot}}$ for a covering factor of unity. 
By increasing the magnetic field strength, we go from a picture of small dense clumps 
to a more diffuse extended medium, which again does not agree with the observational 
upper limit on the H~{\sc i} mass. 
Moreover, high field strengths ($\rm{> 100~\mu G}$) may be found 
in the cores of star formation, but are unlikely to hold on galaxy-wide scales. 
Models with $B_{mag} > 10^{-4}$~G are therefore discarded and appear 
in the shaded grey area on Fig.~\ref{fig:bmag}. 
The model with $B_{mag}$ of $\rm{10^{-4}~G}$ is noted $\mathscr{C}_{B}$. 
Then, if we consider that all of the [O~{\sc i}]~145~$\mu$m emission 
originates in the PDR, we can consequently scale the [C~{\sc ii}] intensity 
and we estimate the contribution of the PDR to the [C~{\sc ii}] line to be 
from 10\% up to at most 50\% in strong magnetic field cases.

\subsection{Origin of the [O~{\sc iii}]~88~$\mu$m line}
\label{sect:oiii}
Although emission from the [O~{\sc iii}]~88~$\mu$m line is expected 
in starburst galaxies, recent {\it Herschel} observations show that it is 
exceptionally bright in dwarf galaxies \citep{madden-2012b}. 
In Haro\,11, it is the brightest of all MIR and FIR lines. 
With an excitation potential of 35.1~eV and critical density, 
n$_{crit} {\rm= 5 \times 10^{2}~cm^{-3}}$, this line is expected to originate 
from a high ionisation and relatively low density medium. 
Effective temperatures, T$\rm{_{eff} > 40~000}$~K, are usually needed 
to explain the bright [O~{\sc iii}] emission in star-forming irregular 
galaxies \citep{hunter-2001}. 
When considering both models $\mathscr{C}_{HII}$ and $\mathscr{D}$, 
we find a good match between model predictions and observations for 
all ionic lines considered, although the [O~{\sc iii}]~88~$\mu$m line is 
under-predicted by the models by a factor of $\sim$2. 
$\mathscr{C}_{HII}$ accounts for 60\% of the [O~{\sc iii}]~88~$\mu$m 
emission, and $\mathscr{D}$ does not predict any [O~{\sc iii}] emission; 
the ionising photons are not energetic enough. 
This factor of 2 is within the uncertainties of this study, however we 
discuss the possible origin of the [O~{\sc iii}]~88~$\mu$m line 
from a highly ionised more diffuse medium. Although the [O~{\sc iii}] emission 
is explained by $\mathscr{C}_{HII}$ in the extreme case of Haro\,11, 
this may not be the case in other metal-poor dwarf galaxies. 

In the star-forming region N11-B of the Large Magellanic Cloud, 
\cite{lebouteiller-2012} find bright [O~{\sc iii}]~88~$\mu$m emission 
that originates from spatially extended high-excitation diffuse ionised gas, 
and can be modeled by O stars distributed across the region.
In the case of Haro\,11, the density in $\mathscr{C}_{HII}$ 
is too high to fully reproduce the observed [O~{\sc iii}]~88~$\mu$m emission. 
As shown in Fig.~\ref{fig:hiidiag}, lowering the density in 
$\mathscr{C}_{HII}$ to densities $\le$ 10$^{2}$ cm$^{-3}$ 
would better reproduce the observed [O~{\sc iii}]~88~$\mu$m intensity. 

Given the two components that we have modeled $\mathscr{C}_{HII}$ 
and $\mathscr{D}$, we may consider 
a picture in which $\mathscr{C}_{HII}$ does not have a covering 
factor of unity, but is actually porous. 
This would allow some fraction of the ionising photons to escape the compact 
H~{\sc ii} region and travel further away in a lower density medium. 
Such a configuration could explain the [O~{\sc iii}]~88~$\mu$m emission, 
but would also lead to over-predicting several lines. 
In particular, in our case, predicted intensities of the [S~{\sc iv}], [O~{\sc iv}], and [N~{\sc iii}] 
lines would be a factor of $\sim$2 higher than their observed values.

\subsection{Origin of the [O~{\sc iv}] line and X-rays}
\label{sect:xrays}
With an ionisation potential of 54.9~eV, [O~{\sc iv}] can only be excited 
by high energy sources. 
Its origin has been discussed in \cite{lutz-1998,schaerer-1999}. 
It can come from AGN activity (X-rays), very hot sources or shocks. 
In this Section, we explore the role of X-ray photons and/or shocks.
In dwarf starbursting galaxies, a hot and young stellar population is usually 
responsible for the [O~{\sc iv}] emission. This has been confirmed by ISO 
observations of the dwarf galaxies II\,Zw\,40 and NGC\,5253 \citep{schaerer-1999}.
Signatures of Wolf-Rayet stars in the optical and P Cygni profiles 
in the far-UV (O~{\sc vi} lines), revealing the presence of O supergiants 
and a very young burst in Haro\,11, were reported by \cite{bergvall-2006} 
and \cite{grimes-2007}.
In the case of Haro\,11, we find that the young starburst containing Wolf-Rayet stars 
can indeed account for the [O~{\sc iv}]~25.9~$\mu$m line intensity. 
The luminosity of the [O~{\sc iv}] line is $\sim$0.006\% of L$\rm{_{TIR}}$. 
\cite{spinoglio-2012} have compared the IR/submm emission lines to 
the IR luminosity in local samples of AGNs and starburst galaxies. 
They showed that the [O~{\sc iv}] line is weaker in starburst galaxies, 
about one order of magnitude less intense than in AGNs, and that starbursts 
follow the relation $\rm{L_{[O~IV]} \sim 10^{-4}~L_{TIR}}$ in the high IR 
luminosity range, which is consistent with that observed in Haro\,11. 
The [Ne~{\sc v}] (97.1~eV) at 14.3~$\mu$m is not detected in the {\it Spitzer}/IRS spectrum. 
The upper limit in Table~\ref{table:lines} is consistent with our model prediction. 
Its absence confirms the fact that there is no AGN activity nor intense 
hard X-ray emission in Haro\,11. 
The X-ray component discussed in Sect.~\ref{sect:irf} in our model 
has little effect on the intensities of the ionic lines.
Around the parameters of $\mathscr{C}_{HII}$, adding X-rays to the radiation field 
increases the intensity of the [O~{\sc iv}]~25.9~$\mu$m line by only $\sim$1\%. 
In the PDR, the X-rays have a moderate effect. 
They contribute up to 5\% to the [C~{\sc ii}] intensity, and $\sim$10\% to the [O~{\sc i}] intensities. 
X-rays play a second order role and the physics of 
the neutral gas is dominated by the FUV photons (PDR), not the X-rays (XDR).

Shocks are also not our prefered explanation for the [O~{\sc iv}] emission. 
\cite{o'halloran-2008} find that the emission of the [Fe~{\sc ii}] 26.0~$\mu$m line 
in low metallicity BCDs may be shock-derived or can result from a larger abundance 
of iron in the gas-phase as it is less depleted onto dust grains at low metallicity. 
However, the [Fe~{\sc ii}] 26.0~$\mu$m line is barely detected in Haro\,11. 
The observed ratio of [Fe~{\sc iii}]/[Fe~{\sc ii}] is $\sim$0.005, which also 
tends to rule out shocks as a major heating process here.

\section{Multi-phase build-up of Haro\,11}
\label{sect:buildup}
\begin{figure*}
\centering
 \includegraphics[clip,trim=0 -0.5cm 0 0,height=7cm,width=7cm]{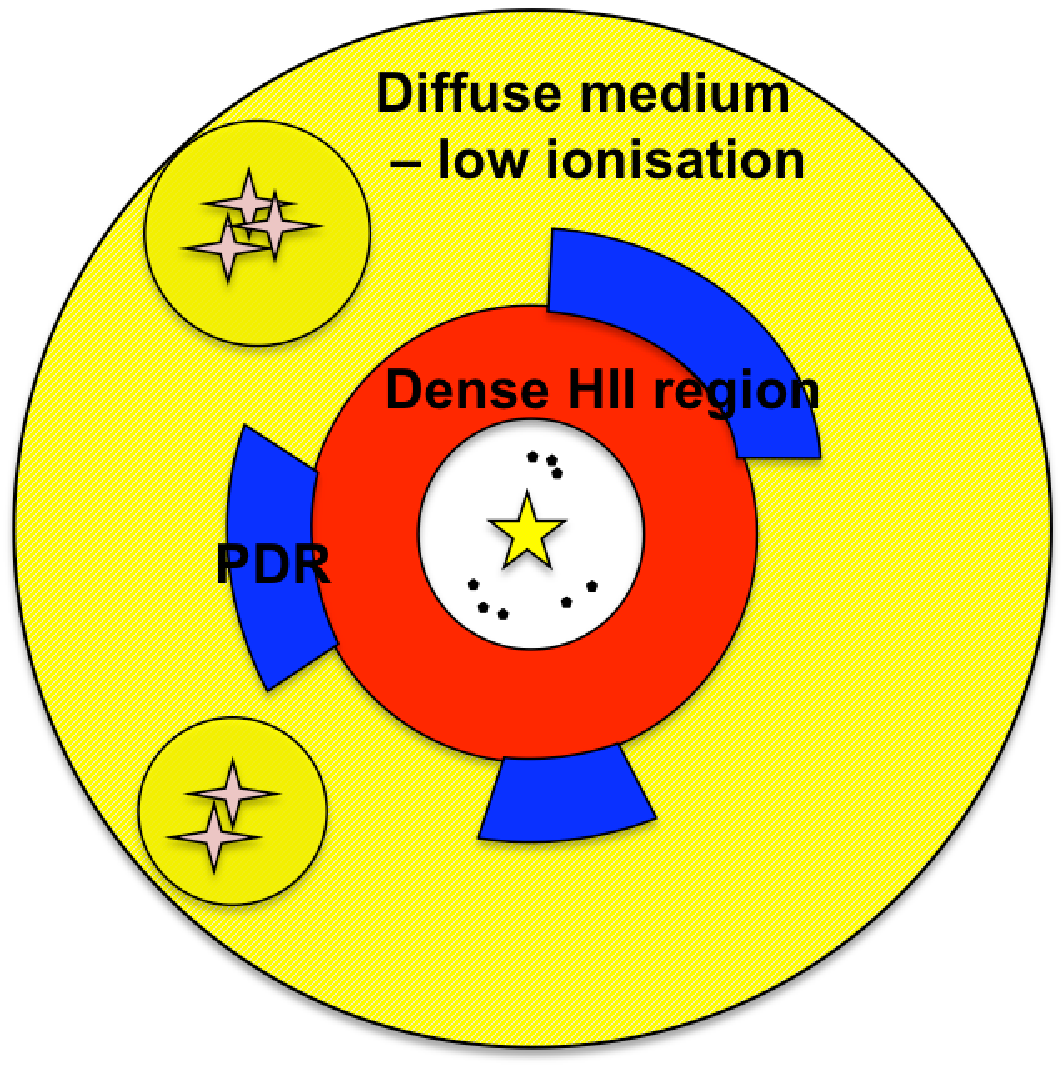}
\includegraphics[clip,height=8cm,width=11cm]{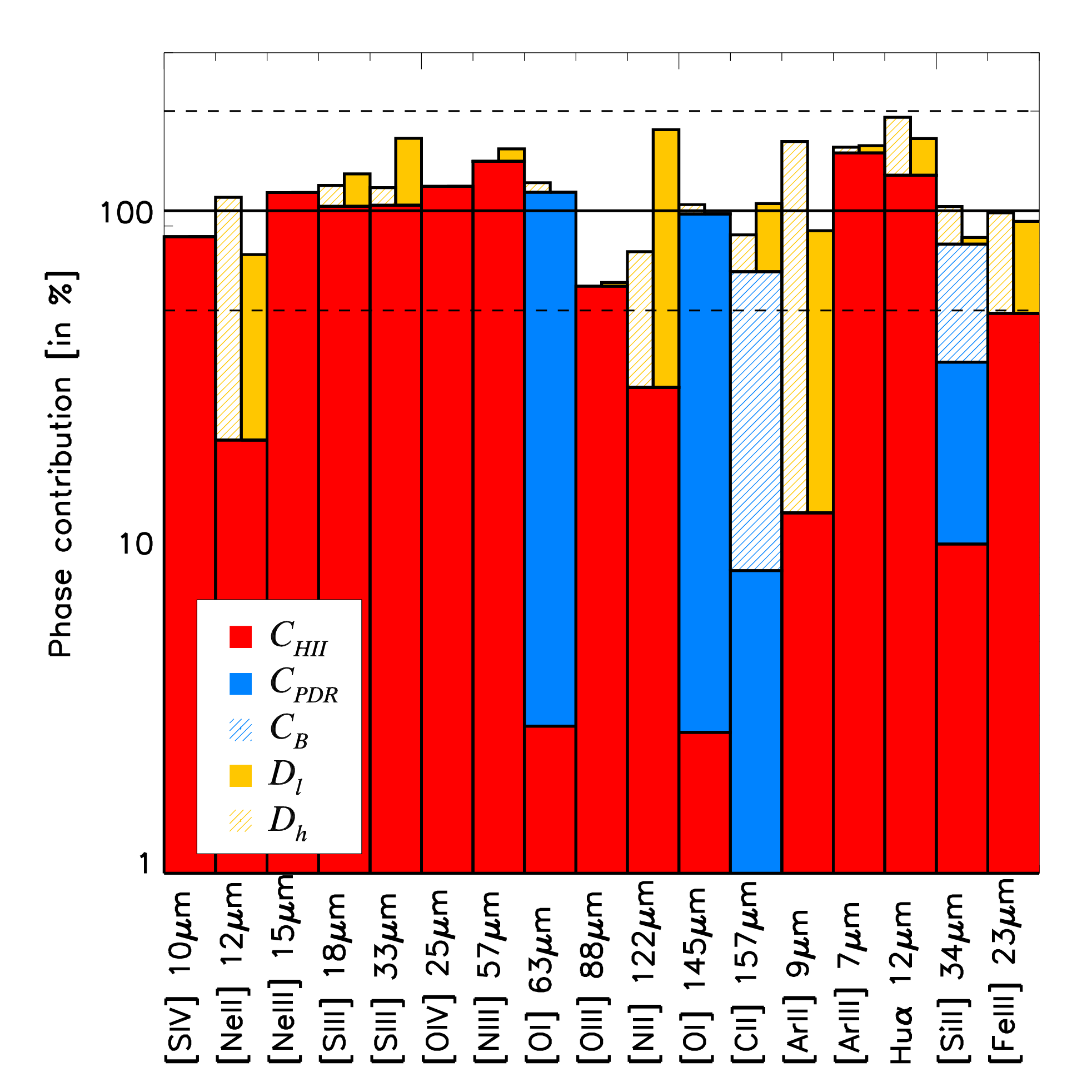}
\caption{
{\it Left:} Multi-phase build-up of the ISM of Haro\,11, 
composed of a compact H~{\sc ii} region ($\mathscr{C}_{HII}$) in red, 
a dense fragmented PDR ($\mathscr{C}_{PDR}$) in blue, 
a diffuse medium of lower ionisation ($\mathscr{D}$) in yellow, 
and warm dust in the inner region (represented as dots close to the starburst).
{\it Right:} 
Diagram of the contribution of each individual ISM phase 
to the global emission of the MIR and FIR lines. 
Red color corresponds to $\mathscr{C}_{HII}$, 
blue to $\mathscr{C}_{PDR}$, with possible enhancement 
in the [C~{\sc ii}] (and [Si~{\sc ii}]) lines when considering 
magnetic fields (hashed blue, $\mathscr{C}_{B}$), 
and yellow from the diffuse ionised medium ($\mathscr{D}_{l}$), 
also with possible contribution from a 
diffuse neutral phase (hashed yellow, $\mathscr{D}_{h}$).
}
\label{fig:multipic}
    \end{figure*}

We summarize the contributions of the different ISM phases that, when 
put together, account for the global MIR and FIR fine-structure line emission 
observed in Haro\,11 (Figure~\ref{fig:multipic}). 
The panel on the {\it left} is an illustration of this multi-phase build-up: 
the central starburst is surrounded by our main model $\mathscr{C}$ composed 
of a compact H~{\sc ii} region $\mathscr{C}_{HII}$ (red), with adjacent dense 
PDRs $\mathscr{C}_{PDR}$ (blue). 
The volume around this compact nucleus is filled by diffuse ionised/neutral gas 
$\mathscr{D}$ (yellow). We also add a low filling factor component of warm dust 
close to the starburst (black dots), the need for which we discuss (Sect.~\ref{sect:sed}). 
This scheme is the global picture (not to scale) 
that results from this modeling study.
We remind the reader that the models 
$\mathscr{C}$ and $\mathscr{D}$ were computed separately. 
The panel on the {\it right} is a histogram indicating the contribution 
of each model to the line intensities, with identical color-coding. 
All 17 lines considered are reproduced by the 3 models $\mathscr{C}_{HII}$, 
$\mathscr{C}_{PDR}$, and $\mathscr{D}$ within a factor of 2. 
The model parameters of these phases and line predictions are 
also listed in Table~\ref{table:sumup}. 
In view of the results from Sect.~\ref{sect:ciidiff} and~\ref{sect:bmag}, in which the 
upper limit on the H~{\sc i} mass limits the contribution from the diffuse neutral medium 
and from magnetic fields, the models $\mathscr{D}_{h}$ and $\mathscr{C}_{B}$ 
are not taken into account for our final build-up of the ISM of Haro\,11. 
The models we combine are: \\
- a compact H~{\sc ii} region that dominates the emission of the ionic lines $\mathscr{C}_{HII}$,\\
- a dense PDR of coverage $\sim$10\% $\mathscr{C}_{PDR}$,\\
- a diffuse low-ionisation medium $\mathscr{D}_{l}$.\\
We analyse results from this build-up in terms of dust emission, extinction, and mass budget.

\begin{center}
\begin{table*}[ht]
  \caption{Model parameters and line predictions.}
  \begin{tabular}{l| l l| l| l| l| l}
    \hline\hline
    \multicolumn{1}{l|}{Quantity} &
    \multicolumn{6}{c}{Model} \\
    \hline
    \multicolumn{1}{l|}{} &
    \multicolumn{1}{l}{$\mathscr{C}_{HII}$~~~~~+} & 
    \multicolumn{1}{l|}{$\mathscr{C}_{PDR}$} &
    \multicolumn{1}{l|}{$\mathscr{D}_{l}$} &
    \multicolumn{1}{l|}{$\mathscr{D}_{h}$} &
    \multicolumn{1}{l|}{Warm dust} &
    \multicolumn{1}{l}{Total~$^{(a)}$} \\
    \hline
	Energy source					& \multicolumn{2}{c|}{Starburst 3.7~Myr} 	& Star 35~000~K	& Star 35~000~K 	& Starburst 3.7~Myr 	&	\\
	n$_{H}$ [cm$^{-3}$]				& 10$^{2.8}$		& 10$^{5.1}$ ( 10$^{3.4}$)	& 10$^{1.0}$	& 10$^{3.0}$	& 10$^{1.5}$ &		\\
	r$_{in}$ [cm]					& 10$^{21.4}$		& 10$^{21.4}$				& 10$^{21.8}$	& 10$^{21.8}$	& 10$^{19.8}$ &	\\
    \hline
	I([S~\sc{iv}]~10.5)/I$_{obs}$~$^{(b)}$	& 0.84 (0.34)	& -	& 0.06	& -		& -	& 0.90	\\
	I([Ne~\sc{ii}]~12.8)/I$_{obs}$		& 0.20 (0.12)	& -	& 0.53	& 0.89	& -	& 0.73	\\
	I([Ne~\sc{iii}]~15.6)/I$_{obs}$		& 1.13 (0.72)	& -	& -		& -		& -	& 1.13	 \\
	I([S~\sc{iii}]~18.7)/I$_{obs}$		& 1.03 (0.59)	& -	& 0.26	& 0.16	& -	& 1.29	 \\
	I([S~\sc{iii}]~33.5)/I$_{obs}$		& 1.04 (0.75)	& -	& 0.61	& 0.14	& -	& 1.65	 \\
	I([Si~\sc{ii}]~34.8)/I$_{obs}$		& 0.11 (0.07)	& 0.24 (0.37)	& 0.04	& 0.24	& - 	& 0.39	\\
	I(Hu~$\alpha$~12.4)/I$_{obs}$		& 1.28 (0.72)	& -	& 0.37	& 0.63	& - 	& 1.65	\\
	I([Fe~\sc{iii}]~22.9)/I$_{obs}$		& 0.49 (0.32)	& -	& 0.44	& 0.50	& -	& 0.93	\\
	I([O~\sc{iv}]~25.9)/I$_{obs}$		& 1.18 (0.87)	& -	& -	& -	& 0.10	& 1.28	 \\
	I([Ar~\sc{iii}]~8.99)/I$_{obs}$		& 1.49 (0.61)	& -	& 0.08	& 0.06	& -	& 1.57	 \\
	I([Ar~\sc{ii}]~6.99)/I$_{obs}$		& 0.12 (0.09)	& -	& 0.75	& 1.49	& -	& 0.87	 \\
	I([C~\sc{ii}]~157)/I$_{obs}$		& 0.01 (0.01)	& 0.07 (0.38)	& 0.40	& 0.19	& -	& 0.48	 \\
	I([O~\sc{iii}]~88)/I$_{obs}$		& 0.59 (0.52)	& -	& 0.02	& -	& -	& 0.61	 \\
	I([O~\sc{i}]~63)/I$_{obs}$			& 0.03 (0.02)	& 1.11 (1.02)	& -	& 0.08	& -	& 1.14	 \\
	I([O~\sc{i}]~145)/I$_{obs}$		& 0.03 (0.02)	& 0.95 (0.98)	& -	& 0.07	& -	& 0.98	 \\
	I([N~\sc{iii}]~57)/I$_{obs}$		& 1.41 (1.17)	& -	& 0.13	& -	& -	& 1.44	 \\
	I([N~\sc{ii}]~122)/I$_{obs}$		& 0.29 (0.28)	& -	& 1.46	& 0.46	& -	& 1.75	 \\
    \hline \hline
  \end{tabular}
  \hfill{}
  \newline
  $(a)$ The total is taken as the sum of models $\mathscr{C}$, $\mathscr{D}_{l}$ 
  and warm dust, which is what we consider for the final build-up in Sect.~\ref{sect:buildup}; 
  although $\mathscr{D}_{h}$ and $\mathscr{C}_{B}$ may also alter the line predictions.\\
  $(b)$ Ratio of the predicted intensity of the model over the observed intensity (I$_{obs}$). 
  Ratio values below 1\% are not indicated.
  \label{table:sumup}
\end{table*}
\end{center}

\subsection{The global SED}
\label{sect:sed}
The proposed geometry also implies a prediction for the dust 
emission coming from the different ISM phases of Haro\,11. 
We compare the predicted SED of the multi-phase build-up 
with the observations of Haro\,11 in Figure~\ref{fig:sed}. 
The goal here is not to fit in detail the dust emission but rather 
to reproduce the global shape of the SED because it is sensitive 
to the temperature of the dust, which is one of the main 
predictions of the multi-phase model.
The SED of the compact model $\mathscr{C}$ (both $\mathscr{C}_{HII}$ and 
$\mathscr{C}_{PDR}$) peaks at $\sim$40~$\mu$m (Fig.~\ref{fig:sed}), 
as in the observed SED. 
In particular, $\mathscr{C}_{PDR}$ is too dense 
to explain the FIR shape of the SED, which corroborates the presence 
of a more diffuse component. 
Indeed, the diffuse component $\mathscr{D}_{l}$ matches the temperature 
of Haro\,11 in the FIR, although the longest wavelengths, beyond 250~$\mu$m, 
are still under-predicted. 
Perhaps the most surprising result is that the multi-phase model fails 
completely to reproduce the UV-optical extinction and the MIR emission, 
despite the choice of a lower cut in the grain sizes (Sect.~\ref{sect:dustprop}). 
We discuss these two points in Sect.~\ref{sect:uvopt} and Sect.~\ref{sect:sedmir}.

\subsubsection{Energy balance}
\label{sect:uvopt}
In the UV-optical wavelength range, the total outward luminosity 
(Fig.~\ref{fig:sed}: bottom panel, black line) is 80\% greater than 
the observed luminosity. Has the input energy in the UV-optical 
wavelength range been overestimated? We are confident about 
the shape of the input starburst spectrum since the spectral line ratios 
depend on the hardness of the radiation field and are well 
fitted by our models. Additional confidence draws from the right 
amount of input energy as the level of the NIR photometry, unaffected 
by the little extinction present in Haro\,11, is well matched.

The SED reflects the global energy balance of Haro\,11 and shows that 
a fraction of the total input energy injected in the UV-optical wavelength 
range is not absorbed. 
The models $\mathscr{C}_{HII}$ and $\mathscr{D}_{l}$ 
are responsible for the over-prediction in the UV-optical. 
We find the average extinction of the model to be A$_{V} = 0.15$~mag. 
The spectrum is dominated by emission from the ionised model components: 
$\mathscr{C}_{HII}$, of which 90\% of the radiation escapes since 
the PDR covers only 10\%, and $\mathscr{D}_{l}$. 
These models are computed separately and work as MIR-FIR emission line 
diagnostics, very little affected by dust attenuation, but we neglect their interaction. 
We can estimate the reddening needed to match the level of the model 
spectrum to the observed data points. Using the extinction law of 
\cite{calzetti-2000}, with R$_{V} = 3.3$ set by the model, 
we find A$_{V} = 0.9$~mag, hence A$_{B} = 1.2$~mag 
and $\rm{E(B-V) = 0.3}$~mag. 
The reddened spectrum appears in dashed dark line 
on Fig.~\ref{fig:sed} ({\it bottom}). 
From UV continuum measurements and H$\alpha$/H$\beta$ ratio, 
\cite{bergvall-2002} estimate A$_{B} = 0.7$~mag. 
The optical image (Fig.~\ref{fig:acs}) shows a dust lane passing 
in front of knot~B, where the extinction is known to be higher 
\citep{adamo-2010,guseva-2012}. \cite{adamo-2010} find that 
$\rm{E(B-V)}$ is 0.4 towards knot~B and 0.1 towards knot~C.

On the contrary, in the MIR (5 to 40~$\mu$m), the predicted luminosity is twice 
lower than the observed luminosity. This emission in the MIR can be 
attributed to warm dust that we model in the following Section.

This suggests an interpretation in terms of geometry of the galaxy, 
in which a large number of ionising photons can travel far, but 
in the end, the different media (compact and diffuse) are well mixed together, 
yielding a significant dust column along most lines of sight. 
In particular, the warm dust component that we add close to the central 
starburst and emits in the MIR is likely responsible for the absorption 
of the UV-optical part of the $\mathscr{C}_{HII}$ spectrum which is 
the primary contributor to the total spectrum in this range.

\subsubsection{Warm dust component}
\label{sect:sedmir}
\begin{figure*}[!htp]
\centering
\includegraphics[width=18cm]{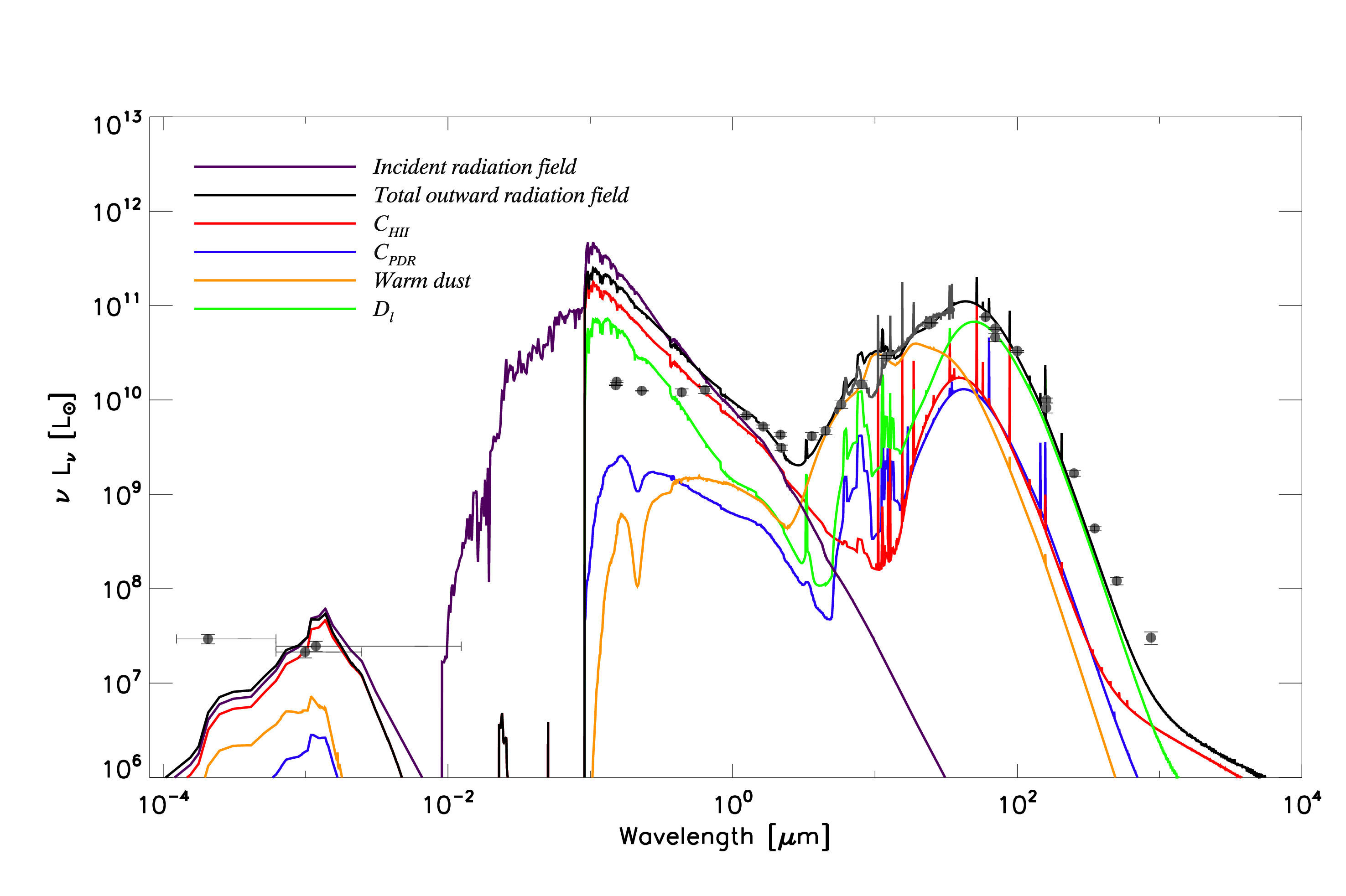}
\includegraphics[width=18cm]{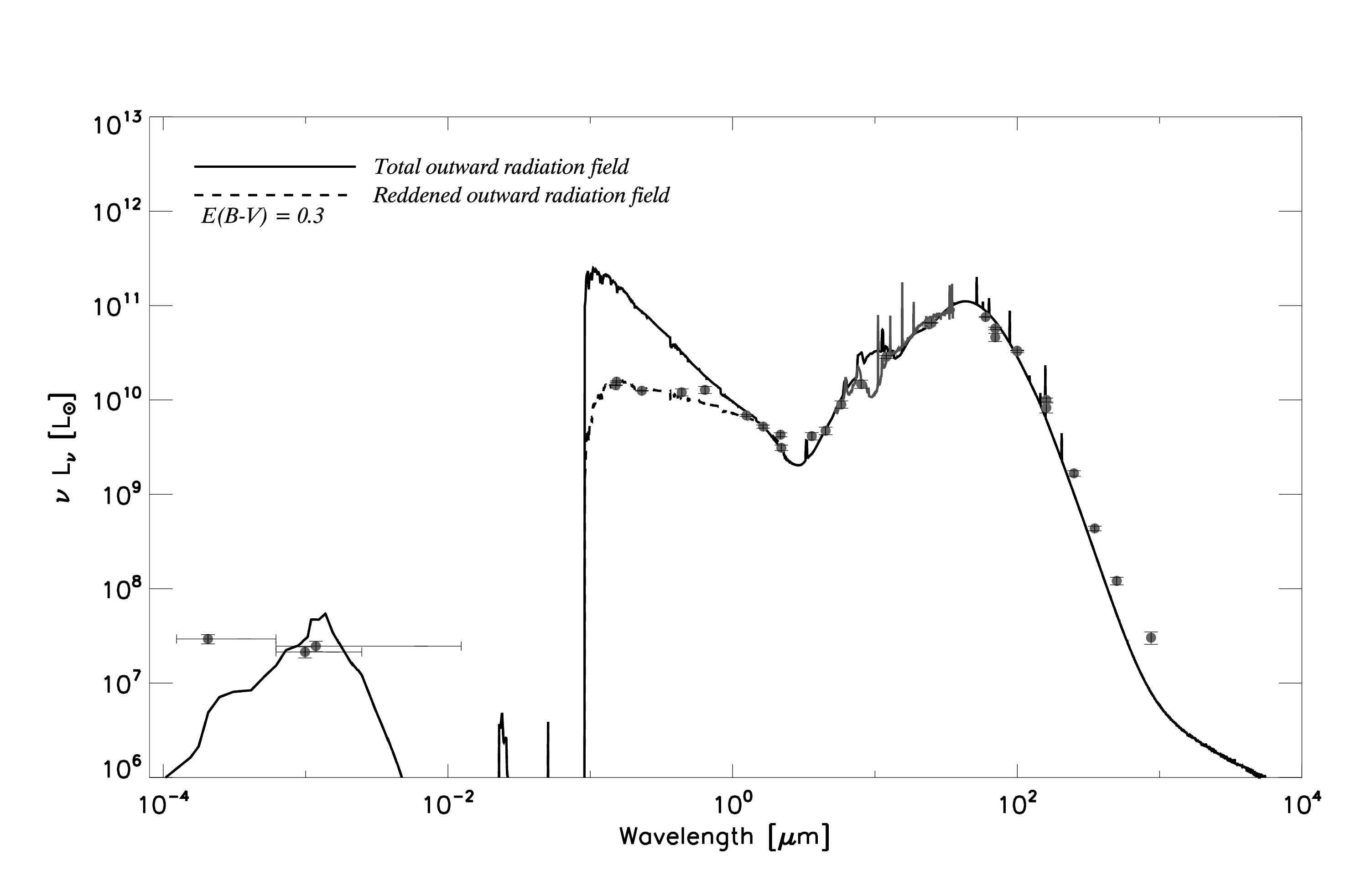}
\caption{
\textit{Top: }
Spectral energy distribution (SED) of the incident radiation field striking the cloud 
and outward radiation from each component modeled by Cloudy. 
The dark spectrum is the total outward radiation resulting from 
multi-phase build-up, and arriving to the observer. 
Photometry data from NED, \cite{bendo-2012}, \cite{galametz-2009}, 
and \cite{remy-2012}, and the IRS spectrum are overlaid in grey.
\textit{Bottom:}
 Spectral energy distribution of the total outward radiation field 
and the reddened outward radiation field in dashed dark line. 
}
\label{fig:sed}
\end{figure*}

To account for the observed warm dust that our models do not produce 
in the MIR range (Fig~\ref{fig:sed}, {\it top} panel), we have examined 
two possible solutions that do not change the intrinsic properties of 
the dust grains (which are little known): \\
\revised{
1) change the dust distribution for a postshock distribution \citep{jones-1996}, 
where grains are affected by shocks, with preferential destruction of large grains, 
resulting in an enhancement of small grains with respect to large grains. 
This grain shattering results in a distribution with steeper slope and lower cut-off sizes. 
This distribution is successfully applied to model IR SEDs of LIRGs in \cite{dopita-2011}. 
} 
\cite{galliano-2003} used this distribution, that may have been generated 
by supernova shocks, to model the dust in the dwarf galaxy NGC~1569. 
However we have been unable to satisfactorily fit the SED by 
changing the grain properties. \\
2) add a warm dust component close to the central starburst. 
In this way we add to our picture a porous layer of dust particles close to 
the central starburst, which is not inconceivable for compact H~{\sc ii} regions. 
Models from \cite{groves-2008} show that the observed warm dust emission 
arises from hot dust embedded within the H~{\sc ii} region. 
This is also supported by spatially resolved observations of nearby galaxies. 
\cite{calzetti-2005} find that the 8~$\mu$m and 24~$\mu$m emission 
originate within the H~{\sc ii} regions.

We adopt a warm dust component, at a distance of $\rm{r_{in} = 10^{19.8}}$~cm, 
with density $\rm{n_{H} = 10^{1.5}~cm^{-3}}$ and covering factor 15\% 
(orange spectrum on Fig~\ref{fig:sed}). These parameters are chosen to 
reproduce the level of the SED in the MIR. In these conditions grains 
can easily survive. 
{\it This component has little or no effect on the discussed emission lines} 
because of its compactness. The photons are predominantly absorbed 
by the dust, not the gas, and dust collisions are responsible for the cooling. 
This warm dust component may be regarded as an equivalent of 
an ultra-compact H~{\sc ii} region \citep{dopita-2003}, or the hot spots 
described in \cite{siebenmorgen-2007}. 
This component takes away 15\% of the input luminosity from the 
H~{\sc ii} region model. The amount of necessary energy and the hardness 
of the radiation field can be adjusted by modifying simultaneously 
the input luminosity, covering factor, and the inner radius in the code. 
Integrating the spectrum from 3 to 1100~$\mu$m gives 
$\rm{L_{TIR} = 1.7 \times 10^{11}~L_{\odot}}$, very close to 
the infered dust-model value of $\rm{1.4 \times 10^{11}~L_{\odot}}$ from 
\cite{galametz-2009,remy-2012}. 

This warm dust model matches the continuum level in the NIR-MIR, 
but produces a silicate feature in emission at 9.7~$\mu$m, 
which is not seen in the {\it Spitzer}/IRS spectrum. 
This discrepancy is likely due to the fact that we have modeled the 
compact and diffuse phases of the ISM independently. 
Foreground dust from the other components may in turn 
extinguish the silicate seen in emission. 
Another factor at play is the adopted abundance 
of silicate grains. Quantitative surveys with {\it Spitzer} of AGB carbon 
and silicate dust production rates in the Magellanic Clouds find that 
carbon dust injection rates dominate over those of silicate 
\citep{matsuura-2009,boyer-2012}. 
There may be a small effect from foreground cold dust 
from the molecular phase of Haro\,11 that we do not model, 
which would have low filling factor and high A$_V$.

\subsubsection{FIR-mm wavelength range}
\label{sect:sedfir}
The mismatch between the observed and predicted SED beyond 250~$\mu$m 
(Fig.~\ref{fig:sed}) may be attributed to colder dust in the molecular phase. 
With no molecular gas constraint, we cannot include a cold molecular 
phase in our modeling. 
However, as a first approximation on the required dust properties, 
we can continue $\mathscr{C}_{PDR}$ to higher A$_V$ and 
stop the model when it matches the submm temperature. 
This way, we estimate that the missing emission requires high A$_V$$\sim$40~mag 
and agrees with a grain temperature of $\sim$25~K, which would fit 
the SPIRE data to 500~$\mu$m but still under-predict the LABOCA 
870~$\mu$m data (see Fig.~\ref{fig:sedmol} in Appendix~\ref{sect:append}). 
The latter is rather related to the submm excess introduced 
in Sect.~\ref{sect:haro11}. 
This excess is modeled by a cold dust component in \cite{galametz-2009}.

\subsection{Optical lines}
Several optical lines, including H$\alpha$, H$\beta$, [O~{\sc iii}]~5007~\AA, 
have been observed towards the central visual peak in \cite{bergvall-2002}. 
Because of the small aperture of the observations, $\rm{4 ^{\prime\prime} \times 4 ^{\prime\prime}}$, 
an absolute flux comparison cannot be made, except for H$\alpha$ 
which was fully mapped in \cite{ostlin-1999} with the ESO 3.6~m telescope. 
More recent observations were also performed with the {\it VLT}/X-shooter, 
in the wavelength range 3~000-24~000~\AA, by \cite{guseva-2012}. 
In particular, they observed, in a slit of size $\rm{\sim 1 ^{\prime\prime} \times 11^{\prime\prime}}$, 
the H$\alpha$, H$\beta$, [O~{\sc ii}]~3727~\AA, [O~{\sc iii}]~4959 and 5007~\AA, 
[S~{\sc ii}]~6716 and 6731~\AA, [S~{\sc iii}]~9069 and 9535~\AA\ lines, in both 
knots B and C of Haro\,11. From these observations, they determined the metallicity 
of Haro\,11 and element abundances.
Optical lines are affected by dust extinction, and comparison of the model predictions 
to the observations is very dependent on the adopted geometry. 
With Cloudy, we can compare the intrinsic emission predicted by our model 
to the extinction-corrected observations. 

The ratio of the optical lines listed above to H$\beta$ are all well reproduced by 
our Cloudy models, within 40\%, with contribution from both the H~{\sc ii} region 
and the diffuse medium (Table~\ref{table:optlines}). 
The only line which poses a problem is [S~{\sc ii}], 
which is over-estimated by a factor of 3 at 6716~\AA, and 1.5 at 6731~\AA.
The Cloudy ratio of the two [S~{\sc ii}] lines is thus 0.9.
The Cloudy models of \cite{guseva-2012} also over-predict the [S~{\sc ii}] lines 
by a factor of 2, but reproduce their observed ratio of 1.5. 
This ratio is sensitive to the electron density of the medium probed, which in 
their case is a density of 10~cm$^{-3}$. We find that our diffuse ionised model 
with density 10~cm$^{-3}$ does predict a [S~{\sc ii}] line ratio of 1.5.
Concerning H$\alpha$, we compare its absolute flux and find that it emits from both 
the H~{\sc ii} region and the diffuse medium, with a total predicted luminosity of 
$\rm{7 \times 10^8~L_{\odot}}$, close to the observed value of \cite{ostlin-1999}, 
$\rm{L_{H\alpha} = 8 \times 10^8~L_{\odot}}$.

\begin{center}
\begin{table}[ht]
  \caption{Optical line predictions.}
  \begin{tabular}{l| l| l| l| l}
    \hline\hline
    \multicolumn{1}{l|}{Line} &
    \multicolumn{4}{c}{I/I(H$\beta$)} \\
    \hline
    \multicolumn{1}{l|}{} &
    \multicolumn{1}{c|}{Observations$^{(a)}$} & 
    \multicolumn{3}{c}{$\mathscr{C}_{HII}$~ +~ $\mathscr{D}_{l}$ ~~=~~ Total Model$^{(b)}$} \\
    \hline
	$\rm{[}$O~\sc{ii}$\rm{]}$~3727~\AA		& \multicolumn{1}{c|}{2.3}		& ~1.4~		& 3.5		& \multicolumn{1}{c}{2.3}		\\
	H$\beta$~4861~\AA					& \multicolumn{1}{c|}{1.0}		& ~1.0~		& 1.0		& \multicolumn{1}{c}{1.0}		\\
	$\rm{[}$O~\sc{iii}$\rm{]}$~4959~\AA		& \multicolumn{1}{c|}{1.1}		& ~1.7~		& 0.08	& \multicolumn{1}{c}{1.5}		 \\
	$\rm{[}$O~\sc{iii}$\rm{]}$~5007~\AA		& \multicolumn{1}{c|}{3.4}		& ~5.2~		& 0.03	& \multicolumn{1}{c}{4.5}		 \\
	H$\alpha$~6563~\AA				& \multicolumn{1}{c|}{2.9}		& ~2.6~		& 2.0		& \multicolumn{1}{c}{2.9}		 \\
	$\rm{[}$N~\sc{ii}$\rm{]}$~6583~\AA		& \multicolumn{1}{c|}{0.53}	& ~0.32~		& 0.79	& \multicolumn{1}{c}{0.52}	 \\
	$\rm{[}$S~\sc{ii}$\rm{]}$~6716~\AA		& \multicolumn{1}{c|}{0.22}	& ~0.07~		& 0.03	& \multicolumn{1}{c}{0.07}	 \\
	$\rm{[}$S~\sc{ii}$\rm{]}$~6731~\AA		& \multicolumn{1}{c|}{0.12}	& ~0.08~		& 0.02	& \multicolumn{1}{c}{0.08}	 \\
	$\rm{[}$S~\sc{iii}$\rm{]}$~9069~\AA		& \multicolumn{1}{c|}{0.15}	& ~0.17~		& 0.13	& \multicolumn{1}{c}{0.19}	 \\
	$\rm{[}$S~\sc{iii}$\rm{]}$~9532~\AA		& \multicolumn{1}{c|}{0.41}	& ~0.42~		& 0.32	& \multicolumn{1}{c}{0.47}	 \\
    \hline \hline
  \end{tabular}
  \hfill{}
  \newline
  $(a)$ Ratio of the observed line intensity over H$\beta$ from \cite{guseva-2012}.
  The fluxes are extinction-corrected and averaged over the two knots~B and C.~
  $(b)$ Ratio of the intrinsic line intensity over H$\beta$ predicted by Cloudy.
  \label{table:optlines}
\end{table}
\end{center}

\subsection{Mass budget}
The gas mass of each ISM component can be calculated 
using the density profile in the cloud and integrating over the 
volume of each slice inside the cloud. 
We then multiply the gas mass by 1.36 to account for helium, 
and use the D/G mass ratio of the model ($\rm{2.64 \times 10^{-3}}$) 
to derive the dust mass. 
The masses of each component are summarized in Table~\ref{table:mass}. 
We find a mass of ionised gas ($\mathscr{C}_{HII}$+$\mathscr{D}_{l}$) of 
$\rm{5.8 \times 10^{8}~M_{\odot}}$, and a mass of neutral gas 
($\mathscr{C}_{PDR}$+$\mathscr{D}_{l}$) of $\rm{2.4 \times 10^{8}~M_{\odot}}$. 
This neutral gas mass accounts for the H~{\sc i} mass as well as the mass 
of $\rm{H_2}$ formed in PDR envelopes before the CO formation. 
This $\rm{H_2}$ layer not traced by CO is referred as the ``dark gas'' 
\citep{wolfire-2010}. 
The PDR-$\rm{H_2}$ mass is $\rm{\sim 10^{8}~M_{\odot}}$ and comes 
from $\mathscr{C}_{PDR}$, while the H~{\sc i} mass comes essentially 
from the diffuse model $\mathscr{D}_{l}$ with 
M(H~{\sc i}) $\rm{\sim 10^{8}~M_{\odot}}$ fixed by its upper limit. 
As expected, the ionised phase is dominant in Haro\,11.
The ionised gas and PDR masses are similar to that of \cite{bergvall-2000}.
We derive a total dust mass of $\rm{2.1 \times 10^{6}~M_{\odot}}$, comparable 
to the value $\rm{6 \times 10^{6}~M_{\odot}}$ of \cite{galametz-2009}. 

\revised{
These dust and gas mass inventories do not include modeling of the cold CO-cloud. 
The missing FIR emission discussed in Sect~\ref{sect:sedfir} results in a dust mass 
of $\rm{\sim2 \times 10^{6}~M_{\odot}}$, which does not include the observed submm excess. 
When modeling the dust submm excess as cold dust, \cite{galametz-2009} 
find a total dust mass of $\rm{2 \times 10^{7}~M_{\odot}}$, which would result 
in a higher D/G mass ratio ($\sim$1/50) compared to our current gas modeling. 
A significant fraction of ISM mass would therefore not accounted for by our models, 
part of which should sit in the molecular phase that we have not modeled. 
CO(1-0) is to date undetected \citep{bergvall-2000}. 
The state of a reservoir of warm molecular gas is addressed in a 
forthcoming paper \citep{cormier-2012c}. 
We would like to emphasize that the determination of the cold dust mass is very sensitive 
to the model and geometry adopted. Both the cool dust interpretation of the submm excess 
and the model accounting for the missing FIR emission (Sect~\ref{sect:sedfir}) are, without 
further onstaints on the cold phase, ad hoc means to fit the dust SED. 
In particular, the cold dust interpretation of the submm excess has little 
physical support as it is unlikely to encounter such very large columns of (cold shielded) 
dust in dwarf galaxies, and because the submm excess does not seem to correlate 
spatially with the densest ISM phases \citep{galliano-2011}. 
}

\begin{center}
\begin{table}[!htp]
  \caption{Mass budget in solar mass $\rm{M_{\odot}}$.}
  \begin{tabular}{l| l l l}
    \hline\hline
    \multicolumn{1}{l|}{Model} & 
    \multicolumn{1}{l}{M(H~\sc{ii})} & 
    \multicolumn{1}{l}{M(PDR)$^{(a)}$} &
    \multicolumn{1}{l}{M$\rm{_{dust}}$} \\
    \hline
    \multicolumn{1}{l|}{$\mathscr{C}_{HII}$} 	& $\rm{3.0 \times 10^{7}}$		& 	-					& $\rm{7.2 \times 10^{4}}$ \\
    \multicolumn{1}{l|}{$\mathscr{C}_{PDR}$} 	& 	-					& $\rm{1.2 \times 10^{8}}$		& $\rm{3.2 \times 10^{5}}$ \\
    \multicolumn{1}{l|}{$\mathscr{D}_{l}$} 		& $\rm{5.5 \times 10^{8}}$		& $\rm{1.2 \times 10^{8}}$ 	& $\rm{1.7 \times 10^{6}}$ \\
    \multicolumn{1}{l|}{Warm dust} 			&	-					& 	-					& $\rm{8.5 \times 10^{3}}$ \\
    \multicolumn{1}{l|}{Total} 				& $\rm{5.8 \times 10^{8}}$		& $\rm{2.4 \times 10^{8}}$		& $\rm{2.1 \times 10^{6}}$ \\
    \multicolumn{1}{l|}{Literature} 			& $\sim \rm{10^{9}}~^{(b)}$	& $\rm{2 \times 10^{8}}~^{(c)}$ 	& $\sim \rm{6 \times 10^{6}}~^{(d)}$ \\
    \hline \hline
  \end{tabular}
  \hfill{}
  \newline
  $(a)$ M(PDR) is the sum of the H~{\sc i} mass and the H$_{\rm 2}$ mass
  in the PDR envelope untraced by CO.~
  $(b)$ \cite{bergvall-2002}.~
  $(c)$ \cite{bergvall-2000}, where they also report M(H~{\sc i}) $\mathrm{< 10^{8}~M_{\odot}}$.~
  $(d)$ This is the dust mass without including the cold component of 10~K from \cite{galametz-2009}. 
    Their total dust mass is $\rm{2 \times 10^{7}~M_{\odot}}$.
  \label{table:mass}
\end{table}
\end{center}

\section{Conclusion}
\label{sect:discussion}
We have presented {\it Spitzer}/IRS and {\it Herschel}/PACS 
spectroscopic observations of the MIR-FIR fine-structure cooling lines 
in the starburst low-metallicity galaxy Haro\,11. 
Investigating the nature of the different gas phases in this galaxy, 
we have modeled the ISM phases of Haro\,11 step by step with the 
spectral synthesis code Cloudy. 
The major model results can be summarized as follows:

\begin{enumerate}

\item
The IRS and PACS spectra show very bright MIR and FIR fine-structure 
cooling lines, the brightest of all being the [O~{\sc iii}]~88~$\mu$m line. 
The galaxy is undergoing young active star formation, as traced by the 
[O~{\sc iii}]~88~$\mu$m line (35.1~eV), which is about 3 times brighter than the classical 
neutral gas tracers [C~{\sc ii}]~157~$\mu$m and [O~{\sc i}]~63~$\mu$m. 
High line ratios of [Ne~{\sc iii}]/[Ne~{\sc ii}], [S~{\sc iv}]/[S~{\sc iii}], and 
[N~{\sc iii}]/[N~{\sc ii}] also trace the hard radiation field. 
We find L$\rm{_{[C~II]}}$/L$\rm{_{TIR}}$~=~0.1\%, 
(L$\rm{_{[C~II]}}$+L$\rm{_{[O~I]}}$)/L$\rm{_{TIR}}$~=~0.2\%, 
and altogether, the MIR and FIR lines represent 1.2\% of L$\rm{_{TIR}}$.

\item
We model our exhaustive dataset of 17 fine-structure lines 
with Cloudy and find the need to describe Haro\,11 with these 
component phases: \\
- a dense H~{\sc ii} region, illuminated by a young starburst, 
from which most of the ionic lines originate, \\
- adjacent dense PDR of low ($\sim$10\%) covering factor, \\
- an extended diffuse low-ionised/neutral medium, \\
- a porous warm dust component close to the stars 
that accountes for the elevated MIR continuum.

\item
Emission from the [N~{\sc ii}]~122~$\mu$m and [Ne~{\sc ii}] lines 
cannot be reconciled with the compact H~{\sc ii} region but are rather 
associated with a more diffuse low-ionisation medium.

\item
We find that 10\% of the [C~{\sc ii}] is associated with the dense PDR, 
and up to 50\% when lowering the density. At least 50\% of the [C~{\sc ii}] 
arises from a more diffuse ionised medium. 
The low extinction and low metallicity of Haro\,11 makes its ISM 
more leaky and the emitting region of [C~{\sc ii}] larger. 
The [O~{\sc i}] emission is fully associated with the PDR. 
The observed [C~{\sc ii}] luminosity is comparable to that of the 
[O~{\sc i}]~63~$\mu$m line. 
In the dense component we derive $\rm{L_{[C~II]}/L_{TIR}\sim0.05\%}$ and 
(L$\rm{_{[C~II]}}$+L$\rm{_{[O~I]}}$)/L$\rm{_{TIR}}$$\sim$1\%, 
which is a common measure of the gas heating efficiency in the neutral gas, 
and in the diffuse component we find $\rm{L_{[C~II]}/L_{TIR}\sim0.1\%}$.

\item
The only line that we do not completely reconcile with our model 
is the [O~{\sc iii}]~88~$\mu$m. Our compact H~{\sc ii} region model 
accounts for 50\% of the emission of [O~{\sc iii}]. 
Our prefered explanation for the intense [O~{\sc iii}] emission 
is the presence of yet another additional diffuse component, filling the volume 
around the porous H~{\sc ii} region, and where the gas is heated by 
escaping UV photons.

\item
We estimate the mass from each modeled phase. 
We find a mass of ionised gas of $\rm{6 \times 10^{8}~M_{\odot}}$, 
a PDR mass of $\rm{2 \times 10^{8}~M_{\odot}}$, and a dust mass 
(without submm constraints, because of the submm excess) 
of $\rm{3 \times 10^{6}~M_{\odot}}$. 
The ionised gas mass is larger than the atomic gas mass.

\item 
Finally, in terms of structure, the ISM of Haro\,11 appears to be mostly 
filled with extended diffuse gas. 
Our simple picture can reproduce the observations by a galaxy with a filling factor 
of diffuse neutral and ionised gas of at least 90\%, a dense H~{\sc ii} region component 
of filling factor $\sim$0.2\% and a PDR component of filling factor $<$0.01\%. 
A radiative transfer model which takes into account clumpy source and 
gas structures would provide a more realistic picture of Haro\,11.

\end{enumerate}

\begin{acknowledgements}
The authors would like to thank Albrecht Poglitsch and Alessandra Contursi 
for their help with the PACS data. 
Part of this work has been made possible by financial support from the CNRS/INSU programme PCMI. 
PACS has been developed by a consortium of institutes led by MPE (Germany) and 
including UVIE (Austria); KU Leuven, CSL, IMEC (Belgium); CEA, LAM (France); 
MPIA (Germany); INAF-IFSI/OAA/OAP/OAT, LENS, SISSA (Italy); IAC (Spain). 
This development has been supported by the funding agencies BMVIT (Austria), 
ESA-PRODEX (Belgium), CEA/CNES (France), DLR (Germany), ASI/INAF (Italy), 
and CICYT/MCYT (Spain). 
  This research has made use of the NASA/IPAC Extragalactic Database (NED) 
  which is operated by the Jet Propulsion Laboratory, California Institute of Technology, 
  under contract with the National Aeronautics and Space Administration. 
Based on observations made with the NASA/ESA Hubble Space Telescope, 
and obtained from the Hubble Legacy Archive, which is a collaboration between 
the Space Telescope Science Institute (STScI/NASA), the Space Telescope 
European Coordinating Facility (ST-ECF/ESA) and the Canadian Astronomy 
Data Centre (CADC/NRC/CSA).

\end{acknowledgements}

\bibliographystyle{aa}
\bibliography{../../BIB/references}

\appendix
\section{SED including the cold molecular component}
\label{sect:append}
\begin{figure*}[!htp]
\centering
\includegraphics[width=18cm]{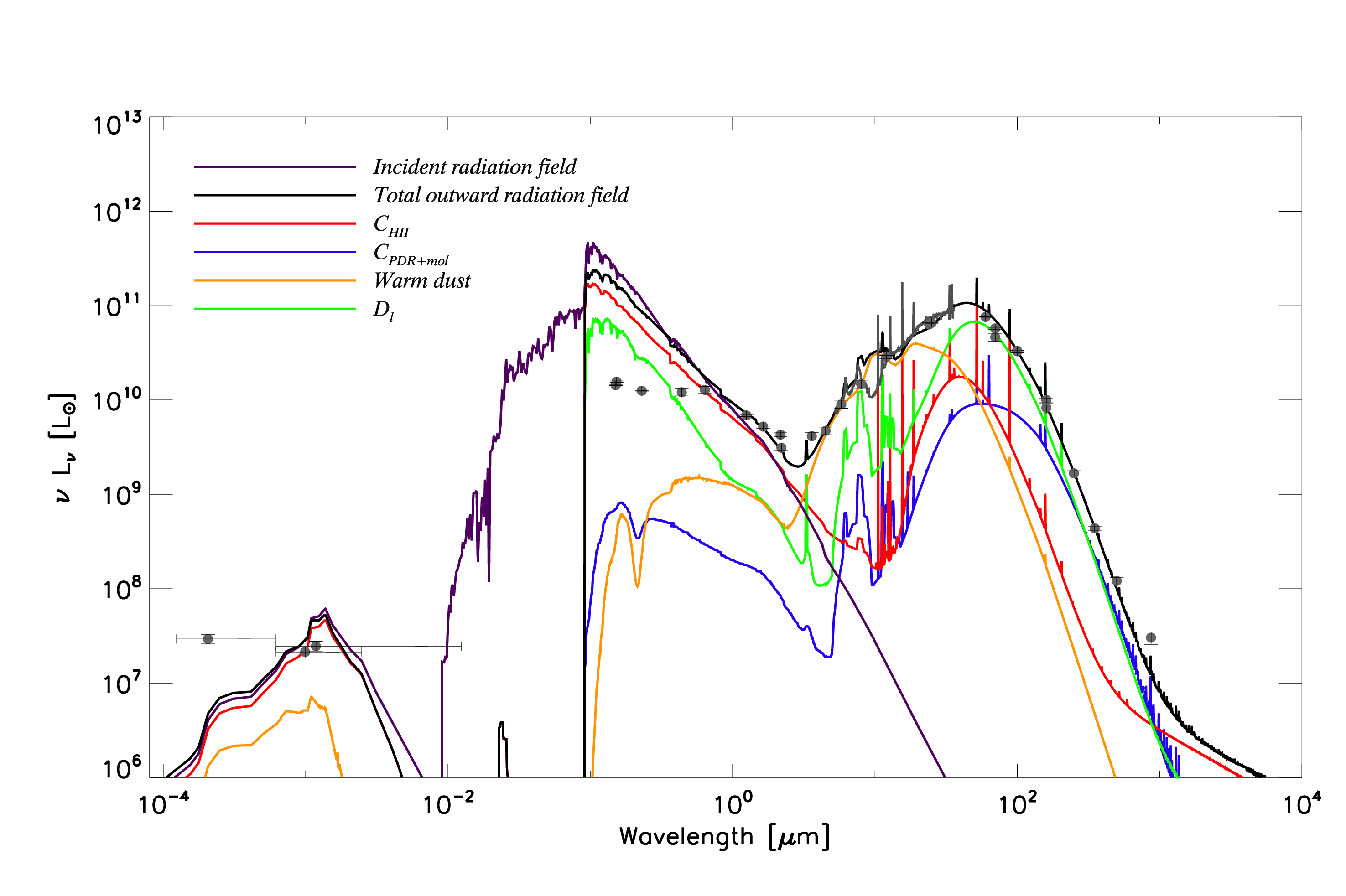}
\caption{
Same as Fig.~\ref{fig:sed} ({\it top} panel) except that the PDR component 
is continued to higher A$_V$ in the molecular regime 
to match the observed submm photometry data. 
The covering factor of this molecular component is of $\sim$5\%.
}
\label{fig:sedmol}
\end{figure*}

We model the SED beyond 250~$\mu$m by stopping the dense PDR model 
$\mathscr{C}_{PDR}$ at larger A$_V$ so that it predicts cold dust emission and 
matches the {\it Herschel}/SPIRE observations (as discussed in Sect.~\ref{sect:sedfir}).
The resulting SED is sown in Fig.~\ref{fig:sedmol}.

\end{document}